\documentclass{aastex63}

\usepackage{amsmath} 
\usepackage{threeparttable}  
\usepackage{tikz}
\usetikzlibrary{shapes,arrows}
\usepackage{float}
\usepackage{multirow} 
\usepackage{tabularx} 

\shorttitle{S-Type stars in LAMOST DR9}
\shortauthors{Chen et al.}
\graphicspath{{./}{figures/}}

\begin{document}

\title{S-type stars discovered in Medium-Resolution Spectra of LAMOST DR9}

\correspondingauthor{A-Li Luo \& Yin-Bi Li}
\email{* lal@nao.cas.cn    ybli@bao.ac.cn}


\author[0000-0001-8869-653X]{Jing Chen}
\affiliation{CAS Key Laboratory of Optical Astronomy, National Astronomical Observatories, Beijing 100101, China}
\affiliation{University of Chinese Academy of Sciences, Beijing 100049, China}

\author[0000-0001-7865-2648]{A-Li Luo$^{*}$}
\affiliation{CAS Key Laboratory of Optical Astronomy, National Astronomical Observatories, Beijing 100101, China}
\affiliation{University of Chinese Academy of Sciences, Beijing 100049, China}
\affiliation{School of Information Management \& Institute for Astronomical Science, Dezhou University, Dezhou 253023, China}

\author[0000-0001-7607-2666]{Yin-Bi Li$^{*}$}
\affiliation{CAS Key Laboratory of Optical Astronomy, National Astronomical Observatories, Beijing 100101, China}
\affiliation{University of Chinese Academy of Sciences, Beijing 100049, China}

\author{Xiang-Lei Chen}
\affiliation{CAS Key Laboratory of Optical Astronomy, National Astronomical Observatories, Beijing 100101, China}
\affiliation{University of Chinese Academy of Sciences, Beijing 100049, China}

\author{Rui Wang}
\affiliation{CAS Key Laboratory of Optical Astronomy, National Astronomical Observatories, Beijing 100101, China}

\author{Shuo Li}
\affiliation{CAS Key Laboratory of Optical Astronomy, National Astronomical Observatories, Beijing 100101, China}
\affiliation{University of Chinese Academy of Sciences, Beijing 100049, China}

\author{Bing Du}
\affiliation{CAS Key Laboratory of Optical Astronomy, National Astronomical Observatories, Beijing 100101, China}

\author[0000-0002-9279-2783]{Xiao-Xiao Ma}
\affiliation{CAS Key Laboratory of Optical Astronomy, National Astronomical Observatories, Beijing 100101, China}
\affiliation{University of Chinese Academy of Sciences, Beijing 100049, China}

\begin{abstract}

In this paper, we report on 606 S-type stars identified from Data Release 9 of the LAMOST medium-resolution spectroscopic (MRS) survey, and 539 of them are reported for the first time. The discovery of these stars is a three-step process, i.e., selecting with the ZrO band indices greater than 0.25, excluding non-S-type stars with the iterative Support Vector Machine method, and finally retaining stars with absolute bolometric magnitude larger than -7.1. The 606 stars are consistent with the distribution of known S-type stars in the color-magnitude diagram. We estimated the C/Os using the [C/Fe] and [O/Fe] provided by APOGEE and the MARCS model for S-type stars, respectively, and the results of the two methods show that C/Os of all stars are larger than 0.5. Both the locations on the color-magnitude diagram and C/Os further verify the nature of our S-type sample.
Investigating the effect of TiO and atmospheric parameters on ZrO with the sample, we found that log g has a more significant impact on ZrO than Teff and [Fe/H], and both TiO and log g may negatively correlate with ZrO. According to the criterion of \cite{2020ApJS..249...22T}, a total of 238 binary candidates were found by the zero-point-calibrated radial velocities from the officially released catalog of LAMOST MRS and the catalog of \cite{2021ApJS..256...14Z}. A catalog of these 606 S-type stars is available from the following link
\href{https://doi.org/10.12149/101097}{https://doi.org/10.12149/101097}.

\end{abstract}

\keywords{stars: late-type --- stars: AGB and post-AGB --- stars: evolution --- methods: data analysis --- techniques: spectroscopic}

\section{Introduction} \label{sec:intro}

S-type stars was originally defined by \cite{1922ApJ....56..457M},
and their spectra show characteristic of s-process element enrichments (especially strong ZrO bands) \citep{1990ApJS...72..387S, 1954ApJ...120..484K}, which are produced through the third dredge-up process in the thermally pulsing asymptotic giant branch (TP-AGB) stage. S-type stars can be classified into two categories, i.e., intrinsic and extrinsic S-type stars. The intrinsic S-type stars are most likely be members of the TP-AGB stars, and have technetium lines in their spectra (Tc-rich S-type stars). The extrinsic S-type stars, without Tc lines in their optical spectra (Tc-poor S-type stars), are in a binary system, and their s-process over-abundances were caused by mass transfer of a former AGB companion which is a white dwarf now \citep{1988A&A...198..187J}.


As a kind of AGB stars, S-type stars are at the intermediate evolutionary stage between M-type giants ($C/O < 0.5$) and carbon stars ($C/O > 1.0$) \citep{1983ARA&A..21..271I}, and the atmospheres of AGB stars are characterized by an increasing enrichment of $^{12}C$ and s-process elements, which increase along the spectral sequence M-MS-S-SC-C.
MS stars are typically at C/O=0.5 to 0.7, S stars are above this range and up to more than 0.95, 
and SC stars represent a rare but important transition group between S stars and carbon stars. Different from M-giants and carbon stars, S-type stars (MS, S, and SC) have not been well studied, and it is undoubtedly important for understanding the evolutionary stage from M giants to carbon stars and helping to clarify the s-processes taking place inside stars.

Catalogues of S-type stars have yielded via searches in survey observations, such as the Henize sample \citep{1960AJ.....65..491H} and the three versions of the General Catalogue of S stars \citep[hereafter, GCSS;][]{1976PWSO...2...21S, 1984PWSO...3....1S, 1990AJ....100..569S}. \citet{1960AJ.....65..491H} provided a sample of 205 S stars located south of $\delta = -25^{\circ}$ and brighter than $R=10.5$, which are inherited in \citet{2000A&AS..145...51V}. \cite{1976PWSO...2...21S} constructed the first version of GCSS including 741 S-type stars, \cite{1984PWSO...3....1S} listed all the 1347 S-type stars whose positions were known at that time, which is nearly twice as large as the first version in \cite{1976PWSO...2...21S}. Furthermore, \cite{1990AJ....100..569S} published the latest version of GCSS including 75 new and fainter S stars located in the Northern Milky Way. The Henize sample was used to investigate the properties of the S stars, such as technetium dichotomy, the proportion of extrinsic and intrinsic S stars, and new symbiotic stars  \citep{1999A&A...345..127V, 2000A&A...360..196V, 2000A&AS..145...51V}. \cite{2018A&A...620A.148S, 2019A&A...625L...1S, 2020A&A...635L...6S} used the GCSS to study the stellar parameters and chemical abundances of S-type stars, and found the evidence of third dredge-up occurrence in S-type stars and the case of bitrinsic S-type stars. \cite{2019AJ....158...22C} used the GCSS to investigate their infrared properties, and found 172 new intrinsic S-type stars.

The atmospheric parameters of S-type stars are difficult to determine, because their atmospheric structure is not only related to the effective temperature, surface gravity and metallicity, but also sometimes is severely impacted by the C/O, and even more with the elements of the s process abundance. However, some previous works still tried to estimate the atmospheric parameter for S-type stars. The (V-K, J-K) and (TiO, ZrO) diagnostics can be used to estimate the atmospheric parameters of S-type stars in \cite{2003IAUS..210P..A2P}, such as $\rm{T_{eff}, C/O, [Fe/H], and [s/Fe]}$. MARCS models were published covering the entire parameter space of S-type stars \citep{2003IAUS..210P..A2P, 2008A&A...486..951G, 2017A&A...601A..10V}, and \citet{2018A&A...620A.148S, 2019A&A...625L...1S, 2020A&A...635L...6S} used these models and HERMES spectra \citep{2011A&A...526A..69R} to calculate element abundances of some S-type stars. 

Most S-type stars in above catalogues are mainly from the southern sky and outside of galactic latitude region $\pm 10^{\circ}$, as a survey of the northern sky, the Large Sky Area Multi-object Fiber Spectroscopic Telescope (LAMOST) can undoubtedly increase the samples of S-type stars, which will provide more information for further studies. In this work, we searched, identified and analyzed S-type stars using the Ninth Data Release (DR9) medium-resolution spectra(MRS) from LAMOST. The paper is organized as follows. The methods for searching S-type stars are presented in Section \ref{sec:data}, and the further identification is introduced in Section \ref{sec:validation}. In Section \ref{sec:discuss}, we discussed the relationship of ZrO vs. TiO and ZrO vs. atmospheric parameters, and selected binary candidates with radial velocities. Finally, the summary are presented in Section \ref{sec:summary}.

\section{Data} \label{sec:data}

\subsection{LAMOST and Gaia} \label{lamost_gaia}

LAMOST is a 4 m Schmidt telescope, which is equipped with 4000 fibers in a field of view of 20 $\rm deg^2$ in the sky \citep{1996ApOpt..35.5155W, 2004ChJAA...4....1S, 2006ChJAA...6..265Z, 2012RAA....12..723Z, 2012RAA....12.1197C, 2012RAA....12.1243L}. Since October 2018, LAMOST started the phase II survey, which contains both low- and medium- resolution spectroscopic surveys. On 2020 Jun, LAMOST delivered its eighth data release (DR8), which provides 17,628,024 single exposure and 4,728,861 co-added MRS spectra. From September 2020 to Jun 2021, LAMOST collected additional 6,054,085 single-exposure spectra and 1,724,868 combined spectra, which will be officially released in March of this year as a part of DR9 \footnote{\href{lamostdr9}{http://www.lamost.org/dr9/}}. For a target of MRS, there are two spectra within one exposure, i.e., a blue and a red band spectra, their wavelength ranges are 4950-5350 $\rm \AA$ and 6300-6800 $\rm \AA$, and they both have a resolution of 7500 respectively at 5163 $\rm \AA$ (blue) and 6593 $\rm \AA$ (red). In this data release, five spectroscopic parameter catalogs are published, the LAMOST MRS General Catalog was used in this work.

Gaia is the space-astrometry mission of the European Space Agency, and the primary science goal of Gaia is to examine the kinematical, dynamical, and chemical structure and evolution of our Milky Way \citep{2016A&A...595A...1G}. The science data of Gaia comprise absolute astrometry (positions, proper motions, and parallaxes), as well as three broad-band photometry ($G$ band, $G_{BP}$ and $G_{RP}$ ) for all objects. Gaia data release 2 (DR2) provides precise positions, proper motions, parallaxes, and photometries for over 1.3 billion stars brighter than magnitude 21 \citep{2018A&A...616A...1G}.

\subsection{Sample Selection of Low-Temperature Giants} \label{sample}

S-type stars are low-temperature giants, and we selected such giant sample from LAMOST DR9 at first. However, there is neither stellar parameters nor stellar types in the LAMOST MRS General Catalog, the low-temperature giants were selected from the color magnitude diagram. As listed in Table \ref{tab:KnownSstar}, we collected 2755 S-type stars from eight catalogs of the literature, which include the four catalogs mentioned in Section \ref{sec:intro}, they were cross-matched with Gaia DR2, and only 1076 stars with $\sigma_{\varpi} / \varpi \leq 0.3$ were retained.

\begin{figure}[ht!] 
	\centering
        \includegraphics[width=0.5\textwidth]{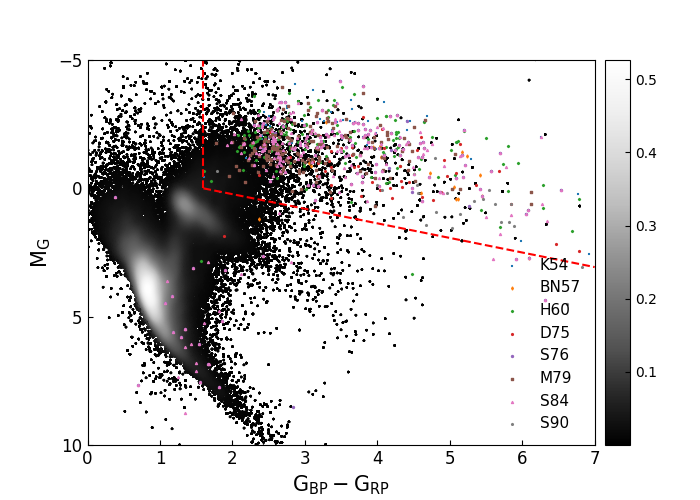}
	\caption{The location of low-temperature giant sample in Hertzsptrung-Russell diagram. The dots with different colors denote the known S-type stars in eight catalogs listed in Table \ref{tab:KnownSstar}, and the red dashed lines indicate the criterion of selecting the low-temperature giant sample. The colour-coded density distribution represents the background common stars of LAMOST and Gaia. \label{LAMOST_HRD}} 
\end{figure}

The distribution of the 1076 stars on the color magnitude diagram of Gaia is shown in Fig.\ref{LAMOST_HRD}, and the absolute magnitude $M_{G}$ was calculated by 
\begin{gather} 
M_{G} = G + 5 - 5log_{10}r -A_{G},
\end{gather}
where G is Gaia G magnitude, r is heliocentric distance, and A$_{G}$ is extinction. r was simply estimated by inverting the parallax, and the parallax zero-point was not considered, which may range from -29 $\mu$as to -80 $\mu$as as mentioned in \cite{2019ApJ...878..136Z}. 
In addition, A$_{G}$ was also not considered because the Gaia individual extinction estimates are rather poor for most stars.

From Fig.\ref{LAMOST_HRD}, we can see that most S-type stars are distributed to the upper right of the red dashed line although a small number of them fall on the main sequence, thus we select the low-temperature giants with the color and magnitude criteria shown by the red dashed lines as follows :
\begin{gather} 
M_{G} > 5 \cdot (G_{BP} - G_{RP}) / 9 -1, \label{two} \\
G_{BP} - G_{RP} \ge 1.6, \label{three}
\end{gather}
where $G_{BP}$ and $G_{RP}$ are the blue and red band magnitudes of Gaia, respectively.

The LAMOST MRS General Catalog was cross-matched with Gaia DR2 using a radius of 3$''$, and 29,040,187 spectra have Gaia data. Using the above criteria in equation \eqref{two} and \eqref{three}, a sample of 1,099,403 spectra of low-temperature giants were selected.

\begin{table*}[ht]
	\centering
	\caption{The number of S-type stars in the literature.}\label{tab:KnownSstar}
	\begin{threeparttable}
	    \begin{tabular}{cccccc} \hline \hline
		    Reference & number \tnote{a} & $\mathrm{Xmatch_{Gaia}}$ \tnote{b}& Gaia Selected \tnote{c}& $\mathrm{Xmatch_{LAMOST}}$ \tnote{d} & LAMOST Selected \tnote{e}\\ \hline
		    \cite{1954ApJ...120..484K} & 87 & 87 & 75 & 10 & 2\\ 
		    \cite{1957ApJ...125..408B} & 29 & 29 & 19 & 3 & 2\\
		    \cite{1960AJ.....65..491H} & 199 & 199 & 184 & 0 & 0\\
		    \cite{1975AbaOB..47....3D} & 73 & 73 & 54 & 29 & 18\\
		    \cite{1976PWSO...2...21S} & 741 & 263 & 178 & 37 & 0\\
		    \cite{1979AAS...38..335M} & 204 & 204 & 189 & 1 & 1\\
		    \cite{1984PWSO...3....1S} & 1347 & 487 & 348 & 31 & 27\\
		    \cite{1990AJ....100..569S} & 75 & 75 & 29 & 2 & 1\\ \hline
	    \end{tabular}
	        \begin{tablenotes}
	        \footnotesize
	            \item{$^a$} The number of S-type stars in each literature.
	            \item{$^b$} The number of S-type stars after matching with Gaia DR2 with 3 $''$.
	            \item{$^c$} The number of S-type stars with high qualities of Gaia parallax.
	            \item{$^d$} The number of S-type stars after matching with LAMOST DR9 with 3 $''$.
	            \item{$^e$} The number of S-type stars with high spectral qualities of LAMOST DR9.
	        \end{tablenotes}
    \end{threeparttable}
\end{table*}

\subsection{Band-strength indices} \label{lineindex}
As in \cite{1954ApJ...120..484K}, the absorption from the ground state of the ZrO molecule gives rise to three band systems, $\alpha$-System in the blue, $\beta$-System in the yellow, and $\gamma$-System in the red. According to the LAMOST MRS wavelength range introduced in Section \ref{lamost_gaia}, the $\gamma$-System in the red band spectra was used, we selected four most intensity ZrO molecular bands of $\gamma$-System according to \cite{1979ApJ...234..538A}, and their wavelength ranges in \cite{2019AcA....69...25G} are slightly changed after considering much higher spectral resolution of LAMOST MRS.

We calculated the ZrO band strengths of S-type stars and M giants to verify whether there is a difference between the two type stars. The band strength indices were computed as follows:
\begin{gather}
B=1-\frac{\lambda_{C,f} - \lambda_{C,i}}{\lambda_{B,f} - \lambda_{B,i}}\frac{\int_{\lambda_{B,i}}^{\lambda_{B,f}} F_{\lambda} d{\lambda}} {\int_{\lambda_{C,i}}^{\lambda_{C,f}} F_{\lambda} d{\lambda}}, \label{four}
\end{gather}
where $F_{\lambda}$ is the observed flux in the wavelength range($\lambda, \lambda + d\lambda$), and the continuum window $\lambda_{C,f} - \lambda_{C,i}$ and the band window $\lambda_{B,f} - \lambda_{B,i}$ are listed in Table \ref{tab:ZrOband}, which are introduced in \cite{2019AcA....69...25G}.

\begin{table}[ht]
	\centering
	\caption{Boundaries of the continuum and band windows of ZrO used in the computation of the band indices.}\label{tab:ZrOband}
	\begin{threeparttable}
		\begin{tabular}{ccccc} \hline \hline
			band & $\lambda_{B,i}$ & $\lambda_{B,f}$ & $\lambda_{C,i}$ & $\lambda_{C,f}$ \\ \hline
			$\mathrm{ZrO_1}$ & 6345.0 & 6352.0 & 6464.0 & 6472.0 \\ 
			$\mathrm{ZrO_2}$ & 6378.0 & 6382.0 & 6464.0 & 6472.0\\
			$\mathrm{ZrO_3}$ & 6474.0 & 6479.0 & 6464.0 & 6472.0\\
			$\mathrm{ZrO_4}$ & 6508.0 & 6512.0 & 6464.0 & 6472.0\\  \hline
		\end{tabular}
	\end{threeparttable}
\end{table}

\begin{figure}[ht!] 
	\centering
        \includegraphics[width=0.5\textwidth]{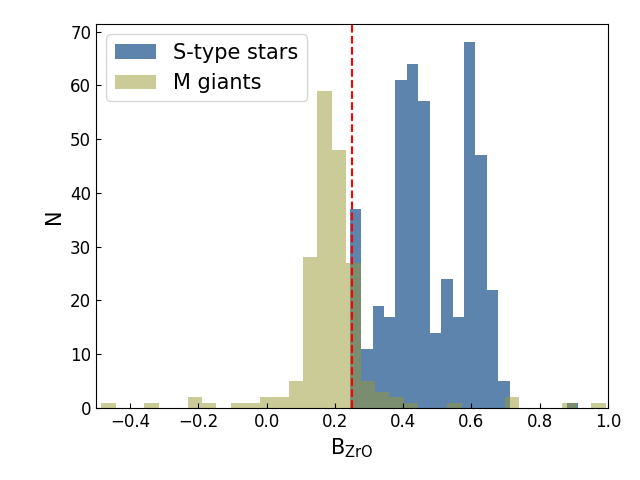}
	\caption{The histogram of the ZrO band indices ($\mathrm{B_{ZrO}}$). The blue and brown histograms represent the $\mathrm{B_{ZrO}}$ distributions for S-type stars and M giants, respectively, and the red dashed line denotes the location of $\mathrm{B_{ZrO} = 0.25}$. \label{ZrO_indices}}
\end{figure}

The spectra of S-type stars were obtained by cross-matching all catalogs in Table \ref{tab:KnownSstar} with LAMOST DR9 MRS, and the common source numbers are listed in the fifth column of Table \ref{tab:KnownSstar}. In addition, only the S-type stars with relatively good quality and obvious ZrO were retained, which are listed in the last column of Table \ref{tab:KnownSstar}. Since there are common sources between different S-type catalogs, we only kept the information in the latest catalog, and the finally selected 460 spectra of 51 stars are listed in Table \ref{tab:51Sstar}. The 200 spectra of M-giants are from S. Li (in preparation, 2022), we excluded the spectra with low signal-to-noise ratio (S/N) and incorrect radial velocities measurements. Using the formula \eqref{four}, we calcualted the ZrO band indices for these S-type stars and M-giants, and the distribution is shown in Fig.\ref{ZrO_indices}.

\begin{table*}[ht]
	\centering
	\footnotesize
	\caption{Basic information of the 51 S-type stars.}\label{tab:51Sstar}
	\begin{threeparttable}
	\begin{tabular}{c|ccccc}
        \hline \hline
        Reference & Identifier$^a$ & $\mathrm{Sptype} ^{b}$ & LAMOST designation & ra & dec \\ \hline
        \multirow{2}{*}{\cite{1954ApJ...120..484K}} & HD 286892 & S5.5/1 & J043537.19+124145.6 & 68.904993 & 12.696 \\ 
        & V* V812 Oph & S5+/2.5 & J174131.93+064341.3 & 265.38307 & 6.7281448 \\ \hline
        \multirow{2}{*}{\cite{1957ApJ...125..408B}} & V* LZ Per & S & J033341.59+485941.5 & 53.423305 & 48.994888 \\ 
        & S1* 141 & S & J061549.23+250040.5 & 93.955129 & 25.011262 \\ \hline
        \multirow{18}{*}{\cite{1975AbaOB..47....3D}} & [D75b] Star 9 & S & J021852.44+624813.6 & 34.718531 & 62.803778 \\
        & [D75b] Star 10 & MS & J021916.76+594221.6 & 34.819836 & 59.706 \\ 
        & [D75b] Star 13 & MS & J022444.76+553545.5 & 36.186504 & 55.595997 \\ 
        & [D75b] Star 14 & MS & J022519.33+581610.6 & 36.33057 & 58.26962 \\ 
        & [D75b] Star 28 & MS & J045915.25+473601.1 & 74.813546 & 47.600324 \\
        & [D75b] Star 31 & S & J051139.28+290621.3 & 77.913696 & 29.105927 \\ 
        & [D75b] Star 33 & MS & J052837.34+340228.0 & 82.155593 & 34.041116 \\ 
        & [D75b] Star 36 & MS & J053511.19+374546.0 & 83.796637 & 37.762804 \\ 
        & [D75b] Star 37 & S & J053652.05+290215.5 & 84.216913 & 29.037664 \\ 
        & [D75b] Star 42 & MS & J060237.80+290705.8 & 90.657512 & 29.118293 \\
        & [D75b] Star 43 & MS & J060527.88+222042.3 & 91.366174 & 22.345094 \\ 
        & [D75b] Star 47 & M2-MS & J061009.58+233301.0 & 92.539955 & 23.550281 \\
        & [D75b] Star 48 & M3III-MS & J061034.57+233854.8 & 92.644077 & 23.648562 \\ 
        & [D75b] Star 49 & MS & J061852.52+211216.0 & 94.718843 & 21.204468 \\ 
        & [D75b] Star 51 & S & J063627.02+070033.1 & 99.112595 & 7.009203 \\
        & [D75b] Star 52 & MS & J063903.22+023400.1 & 99.763447 & 2.566697 \\
        & [D75b] Star 53 & S & J064356.36+014459.6 & 100.98486 & 1.7498975 \\ 
        & [D75b] Star 75 & S & J204555.27+363218.6 & 311.48032 & 36.538501 \\ \hline
        \cite{1979AAS...38..335M} & MC79 1-19 & S3:*4 & J071737.44+002047.2 & 109.40603 & 0.3464478 \\ \hline
        \multirow{27}{*}{\cite{1984PWSO...3....1S}} & CSS  28 & S4,6e & J011941.97+723640.8 & 19.9249 & 72.61135 \\ 
        & CSS 63 &  & J024440.21+530546.2 & 41.16757 & 53.09619 \\
        & CSS 73 &  & J032809.69+174046.7 & 52.040411 & 17.679651 \\ 
        & CSS 96 & S6,1 & J042747.65+222120.0 & 66.94858 & 22.35556 \\ 
        & CSS 97 & Swk & J042841.35+215035.1 & 67.172307 & 21.843089 \\
        & CSS 98 & Swk & J042901.55+234648.6 & 67.256488 & 23.780182 \\
        & CSS 115 &  & J045351.76+320435.7 & 73.46568 & 32.0766 \\ 
        & CSS 138 & Swk & J052545.59+322401.4 & 81.439975 & 32.400405 \\ 
        & CSS 142 &  & J053227.32+405843.8 & 83.11387 & 40.978855 \\ 
        & CSS 148 &  & J053817.61+281144.2 & 84.573412 & 28.195633 \\
        & CSS 214 &  & J062352.89+211830.1 & 95.9704 & 21.308366 \\ 
        & CSS 220 &  & J062715.33+180423.6 & 96.813916 & 18.073228 \\ 
        & CSS 216 &  & J062742.91+521547.1 & 96.928826 & 52.263103 \\
        & CSS 225 &  & J063355.57+213204.2 & 98.481546 & 21.534523 \\ 
        & CSS 241 &  & J064024.65+234358.4 & 100.10274 & 23.732894 \\ 
        & CSS 255 &  & J064532.37+064706.2 & 101.38488 & 6.7850816 \\ 
        & CSS 257 & M7wkS & J064646.93+264114.8 & 101.69558 & 26.687463 \\
        & CSS 273 &  & J065425.98+054703.2 & 103.60827 & 5.7842395 \\ 
        & CSS 305 & M3S & J070606.48-005239.5 & 106.52704 & -0.8776633 \\
        & CSS 331 &  & J072029.17+322317.6 & 110.12156 & 32.388246 \\ 
        & CSS 403 & S4/4e & J074918.17+234404.0 & 117.32572 & 23.734467 \\
        & CSS 413 & M4S & J075253.23+343650.8 & 118.2218 & 34.614128 \\
        & CSS  494 & S3/6e & J082142.85+171705.3 & 125.42858 & 17.28483 \\
        & CSS 986 &  & J172448.18+335508.4 & 261.20075 & 33.919 \\ 
        & CSS 1249 &  & J205232.65+562440.6 & 313.13607 & 56.4113 \\ 
        & CSS 1310 & S6/2 & J224954.43+593935.2 & 342.47683 & 59.659802 \\ 
        & CSS 1328 & SwK & J231608.42+285149.2 & 349.03511 & 28.863688 \\ \hline
        \cite{1990AJ....100..569S} & CSS2 75 & S & J234430.41+560658.7 & 356.12675 & 56.11632 \\ \hline
    \end{tabular}
    \begin{tablenotes}
	    \footnotesize
	    \item{$^a$} The identifier from SIMBAD.
	    \item{$^b$} The spectral type from SIMBAD.
	\end{tablenotes}
    \end{threeparttable}
\end{table*}

As shown in Fig.\ref{ZrO_indices}, the ZrO indices of a small fraction of M giants are in the region of S-type stars, we manually checked these spectra, and found that they are M giant spectra with low S/N ratios or weak ZrO (but the calculated line index does indeed fall in this region), or S-type star spectra. Anyway, S-type stars and M-giant stars can be clearly distinguished with the ZrO bands indices, and we used the criterion of $\rm{{B_{ZrO}>0.25}}$ to distinguish S-type stars from M-giants preliminarily. This criterion was applied to the 1,099,403 low-temperature giant spectra, and 21,121 red band spectra of S-type candidates were obtained after removing the blue band spectra.

\subsection{Selecting S-type stars with the iterative SVM method}

Since the number of selected known S-type stars in Table \ref{tab:51Sstar} is not large, we used the Support Vector Machine (SVM) algorithm, a supervised machine learning algorithm for classification and regression \citep{1995ML...20....273}, to classify S stars from LAMOST DR9 MRS. The advantage of SVM is that it is suitable for small samples, and the complexity of calculation depends on the number of support vectors rather than the dimensionality of the sample space.

\begin{figure}[ht!] 
	\centering
        \includegraphics[width=0.5\textwidth]{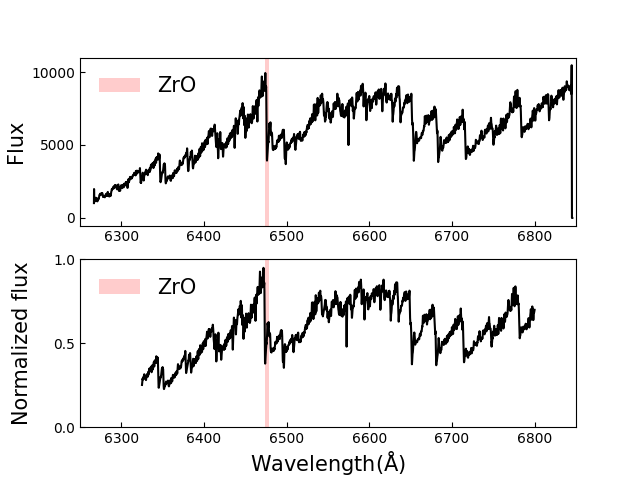}
	\caption{An example before (top) and after (bottom) spectral preprocessing for a red band MRS spectrum. \label{Spectra_process}} 
\end{figure}

Before trainning the SVM classifier, we preprocessed the MRS spectra as the following four steps. (1) In LAMOST FITS file, the ``ormask" flag was set for each wavelength to give six kinds of spectrum problems, and the wavelengths with ``ormask" larger than 0 were removed; (2) Due to the limitation of the wavelength range of the MRS spectra, the ZrO bands are only present in the red band spectra, so the red band spectra were shifted to the vacuum wavelength using the RV released by LAMOST; (3) The spectra were resampled in steps of 0.1 $\rm \AA$ to facilitate subsequent training of the model; (4) S-type stars are low-temperature giant stars with obvious molecular bands in their spectra, removing the pseudo-continuum during normalization may obscure some features, thus we used the maximum flux to normalize them. Figure \ref{Spectra_process} shows an example after spectral preprocessing for a red band MRS spectrum.

In order to train and test the classifier, we need to decide the input features, train and test dataset. The S-type stars are defined with the existence of clearly detectable ZrO bands in visual spectra, thus the previously defined four ZrO bands were adopted as the input features, which are shown in Table \ref{tab:ZrOband}.

There are 51 S-type stars in Table \ref{tab:51Sstar} with LAMOST spectra, which have been observed many times, and their 460 single exposure and coadded red band spectra were treated as the positive samples. We randomly selected the same number of spectra from the low-temperature giant sample as the negative sample, and manually checked the their spectra to avoid the contaminations of S-type stars. The positive and negative samples were randomly divided into training dataset and test dataset according to the ratio of 8:2 to train the SVM classifier and evaluate the performance of the model respectively.

After training and testing, the SVM model was applied to select S-type star spectra from the 21,121 spectra of S-type candidates. The selection process of S-type stars was a gradual iterative process, in which the S-type stars selected in the previous round were used as the positive samples in the next round, the negative samples of the same size in each round were re-selected, and the model was retrained until few new S-type stars can be found.

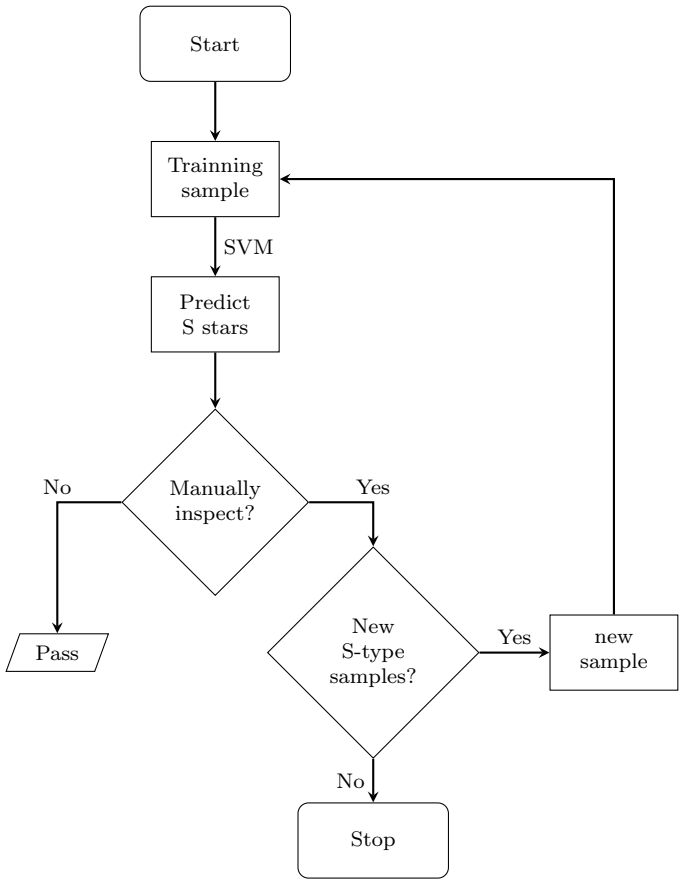
\begin{figure}
    \footnotesize
    \tikzstyle{startstop} = [rectangle, rounded corners, minimum width=2cm, minimum height=1cm,text centered, draw=black, fill=white!30]
    \tikzstyle{io} = [trapezium, trapezium left angle = 70, trapezium right angle=110, text centered, minimum width=0.5cm, minimum height=0.5cm, draw=black, fill=white!30]
    \tikzstyle{process} = [rectangle,minimum width=1.5cm, minimum height=1cm, text centered, text width =1.5cm, draw=black, fill=white!30]
    \tikzstyle{decision} = [diamond, minimum width=1cm, minimum height=1cm, text centered, text width = 1.5cm, draw=black, fill=white!30]
    \tikzstyle{arrow} = [thick,->,>=stealth]

    \begin{tikzpicture}[node distance=2cm]
		\node (start) [startstop] {Start};
		\node (original sample) [process,below of=start,yshift=0.2cm] {Trainning sample};
		\node (predict S stars) [process,below of=original sample, yshift=0.2cm] {Predict S stars};
		\node (manually inspect) [decision,below of=predict S stars,yshift=-0.5cm] {Manually inspect?};
		\node (pass1) [io,left of=manually inspect, xshift=-0.1cm, yshift=-2.0cm] {Pass};
		\node (new S-type sample?) [decision,right of=manually inspect,xshift=0.1cm, yshift=-2.0cm] {New S-type samples?};
		\node (new sample) [process,right of =new S-type sample?,xshift=1.2cm] {new sample};
		\node (stop) [startstop,below of=new S-type sample?, yshift=-0.5cm] {Stop};
		\draw [arrow] (start) -- (original sample);
		\draw [arrow] (original sample) -- node[anchor=west] {SVM} (predict S stars);
		\draw [arrow] (predict S stars) -- (manually inspect);
		\draw [arrow] (manually inspect) -| node[anchor=south] {Yes} (new S-type sample?);
		\draw [arrow] (manually inspect) -| node[anchor=south] {No} (pass1);
		\draw [arrow] (new S-type sample?) -- node[anchor=south] {Yes} (new sample);
		\draw [arrow] (new S-type sample?) -- node[anchor=east] {No} (stop);
		\draw [arrow] (new sample) |- (original sample);
	\end{tikzpicture} 
	\caption{The process of iterative SVM for selecting S-type stars.\label{follow_chart}}
\end{figure}

In the first round, 2060 spectra were predicted as S-type stars, we manually checked these spectra, and 2039 spectra (339 stars) with high spectral quality were left as the positive samples of the second round. In each subsequent round, we all checked whether there were new S-type stars, and put them into the trainning sample. This cycle did not finish until there were only few new S-type stars to be found, and 613 S-type stars were finally selected from LAMOST DR9 MRS after 5 cycles. The specific flow chart of iterative SVM is shown in Fig.\ref{follow_chart}.

\section{Further Investigation} \label{sec:validation}

\subsection{Study the Evolutionary Stage with $M_{bol}$}

From the evolutionary point of view, luminous red stars can be divided into two distinct groups:  red supergiants (RSG) and AGB stars. The supergiants are burning helium (or carbon) in non-degenerate cores, and they are defined as stars with $M \gtrsim 9M_{\odot}$. The AGB stars have completed core burning, and are evolving up the giant branch for the second time burning hydrogen and helium in shells around an electron degenerate carbon/oxygen core. 

It is difficult to distinguish between AGB and RSG because they are both red and luminous, \cite{1983ApJ...272...99W} suggested a criterion to distinguish them by $M_{bol} = -7.1$, which was applied to the 613 S-type stars in this work. We estimated the absolute bolometric magnitude via:
\begin{gather} 
M_{bol} = M_{G} + BC_{G}(T_{eff}), 
\end{gather}
where $M_{G}$ is the absolute magnitude of G band, and $BC_{G}(T_{eff})$ is the temperature-only dependent bolometric correction.  Gaia has defined a function of $BC_{G}(T_{eff})$ in two temperature regions of 2500 - 4000 K and 4000 - 8000 K as follows, with different polynomial coefficients, respectively.
\begin{gather}
BC_{G}(T_{eff}) = \sum_{i=0}^{4} a_{i}(T_{eff} - T_{eff, \odot}),
\end{gather}
where $a_{i}$ is the fitting coefficients, $T_{eff}$ is the temperature of the stars comes from Gaia DR2, and $T_{eff, \odot}$ is solar temperature, which is set to 5772 K \citep{2016AJ....152...41P}. Since the temperature of S-type stars is low, we used the coefficients of 2500 - 4000 K to calculate the bolometric correction. 

\begin{figure}[ht!] 
    \centering
	\includegraphics[width=0.5\textwidth]{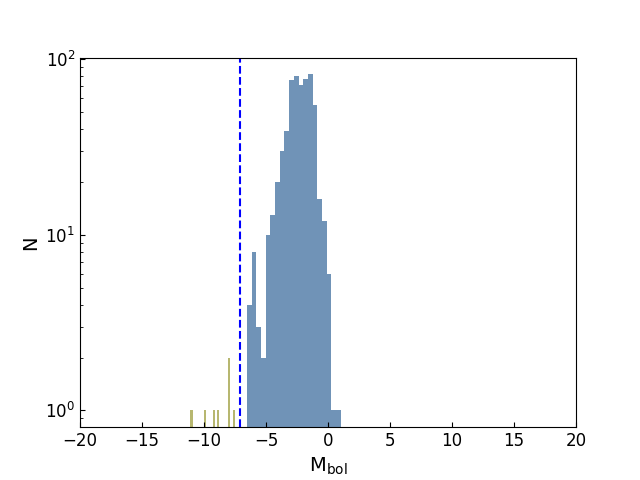} 	
	\caption{The distribution of the absolute bolometric magnitude for 613 S-type stars, and the $\rm M_{bol} = -7.1$ is denoted by the dashed blue line. \label{Bolo_mag}} 
\end{figure}

Figure \ref{Bolo_mag} shows the distribution of the absolute bolometric magnitude for 613 S-type stars, and the dashed line represents $M_{bol} = -7.1$. On the left of the dashed line is the AGB area, the right is the RSG area, and approximately 99.2\% of the 613 S-type stars is in the AGB area. There are 7 sources with $M_{bol} < -7.1$, we crossed them with SIMBAD at 3 arcseconds, and the specific information is shown in the Table \ref{tab:7RSGstar}. Although there are two S-type stars in this table, in order to ensure the reliability of the sample, we still removed the 7 stars from 613 S-type stars, and the remaining 606 S-type stars were the final sample.

\begin{table*}[ht] 	
	\centering 	
	\caption{Basic information for the seven stars with $\mathrm{M_{bol} < -7.1}$.}\label{tab:7RSGstar} 	\begin{threeparttable} 		
		\begin{tabular}{ccccc} 			\hline \hline 			
			LAMOST Designation & RA & DEC & main\_type \tnote{a} & $\mathrm{M_{bol}}$ \tnote{b}\\ \hline 			
			J024933.43+500850.0 & 42.389319 & 50.147238 & - & -7.51\\ 
			J051206.90+454642.7 & 78.028752 & 45.778555 & Mira & -9.13\\
			J062406.62+172350.7 & 96.0276 & 17.397435 & - & -7.98\\
			J071157.54+072958.8 & 107.98979 & 7.4996732 & S* & -8.93\\
			J081035.88+145159.9 & 122.64954 & 14.866662 & - & -7.87\\ 
			J204924.70+362039.8 & 312.35293 & 36.344402 & V* & -9.87 \\
			J234430.41+560658.7 & 356.12675 & 56.11632 & S* & -11.09\\\hline 		
		\end{tabular} 		
		\begin{tablenotes} 			
			\footnotesize 			
			\item $^a$ The main type from SIMBAD.
			\item $^b$ The absolution bolometric magnitude.
		\end{tablenotes} 	
	\end{threeparttable} 
\end{table*}

\subsection{Study the Evolutionary Stage using the CMD}

\cite{2011ApJ...727...53Y} verified 189 RSGs by several CMDs in near- and mid-infrared bands, \cite{2019A&A...629A..91Y} distinguished AGBs and RSGs in the Small Magellanic Cloud with five optical and infrared CMDs, and we studied the evolutionary stage of 613 S-type stars with an infrared CMD to check whether they are in the AGB stage as the theoretically predicted.


Except for 606 S-type stars, known S-type stars, red-giant branch (RGB), RSG, red clump (RC), O-rich AGB (OAGB), C-rich AGB (CAGB) and A, F, G, K, M stars were selected as background stars in the CMD of this work.
The Known S-type stars were from the catalog of \cite{2019AJ....158...22C}, which contains 151 extrinsic and 190 intrinsic S-type stars, and the RGBs were collected from \cite{2018MNRAS.475.3633W}, which has 3726 stars with small age and mass uncertainties. OAGB and CAGB stars were from Table 9 and Table 10 of \cite{2021ApJS..256...43S}, an RC sample of 92249 stars were avaliable from \cite{2018ApJ...858L...7T}, which has a contamination of only ~3\%, and 3325 RSGs of \cite{2011AJ....142..103B} were collected here.
In addition, we also used the A, F, G and K Stars from ``LAMOST LRS Stellar Parameter Catalog of A, F, G and K Stars", the sample of M giants and M dwarfs were obtained from S. Li et al. (in preparation). The 613 S-type stars and all background samples were cross-matched with 2MASS and Gaia DR2 to obtain $J$ and $K$ magnitudes and parallax, respectively. Table \ref{tab:HR_stars} lists the reference, and the numbers before and after the cross-matching.

\begin{table*}[ht]
	\centering 
	\caption{Samples of Color-Magnitude Diagram.}\label{tab:HR_stars}	
	\begin{threeparttable}
		\begin{tabular}{lcccc} 			\hline \hline 			
			Class & Reference & Number \tnote{a} & Selected number \tnote{b}\\ \hline 			
			Predicted S-type stars & This work & 606 & 606 \\ 			
			Known S-type stars & \cite{2019AJ....158...22C} & 341 & 285 \\ 
			Red Giant Branch & \cite{2018MNRAS.475.3633W} & 3,726 & 3,726 \\
			Red Clump & \cite{2018ApJ...858L...7T} & 92,249 & 50,860 \\
			Red Supergiant & \cite{2011AJ....142..103B} & 3,325 & 3,296 \\
			O-rich AGB & \cite{2021ApJS..256...43S} & 5,908 & 3,329 \\
			C-rich AGB & \cite{2021ApJS..256...43S} & 3,596 & 2,747 \\
			LAMOST AFGK & \href{http://dr7.lamost.org/v2.0/catalogue}{http://dr7.lamost.org/v2.0/catalogue} & 6,179,327 & 50,000 \\
			M giant and M dwarf & S. Li (in preparation) & 77,028 & 68,514\\ \hline 		
		\end{tabular} 	
		\begin{tablenotes} 			
			\footnotesize 			
			\item $^a$ The number of each sample in the reference.
			\item $^b$ The number after cross-matching with 2MASS and Gaia DR2.
		\end{tablenotes} 
    \end{threeparttable}
\end{table*}

\begin{figure*}[ht]
	\centering
        \includegraphics[width=0.45\textwidth]{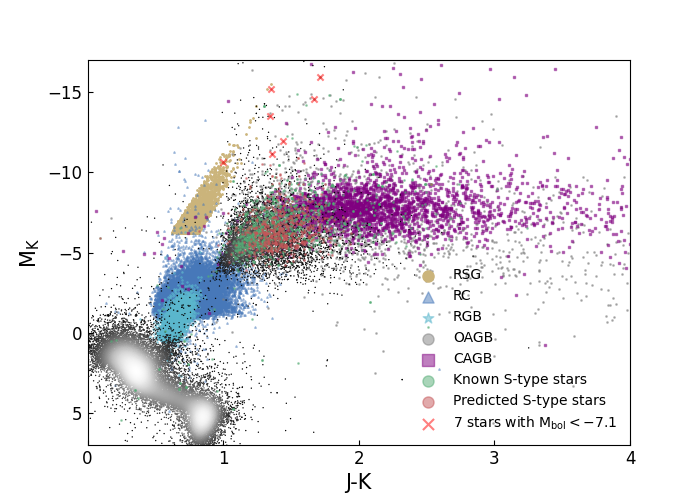}
        \includegraphics[width=0.45\textwidth]{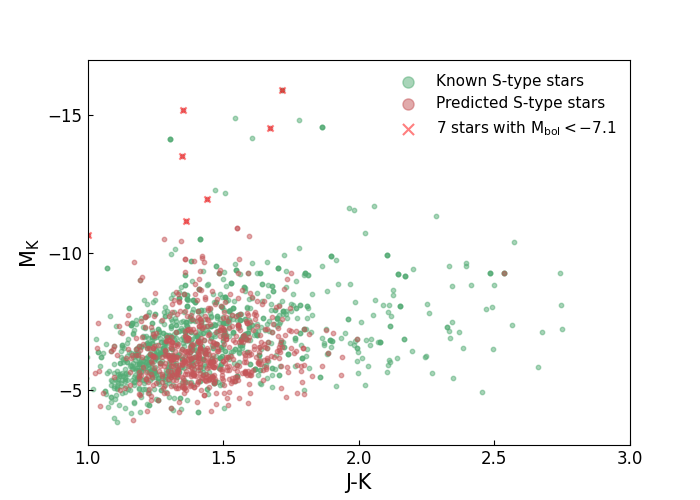}
	\caption{$\rm M_{K}$ vs. J-K color-magnitude diagram for the samples in Table \ref{tab:HR_stars}. The yellow dots, blue triangles, blue stars, grey dots, purple squares, green dots, red dots and red crosses are red super giants, red clumps, red giant branch stars, O-rich asymptotic giant branch stars, C-rich asymptotic giant branch stars, predicted and known S-type stars, and seven stars with $\mathrm{M_{bol} < -7.1}$, respectively, and the colour-coded density distribution denotes the LAMOST A, F, G, K, and M stars. The right panel is an enlargement of the known and predicted S-type star region in the left panel. \label{HR_Diagram}} 
\end{figure*}

The left panel of Fig.\ref{HR_Diagram} shows the CMD of multiple types of stars, the right panel is an enlargement of the AGB region in the left panel. and we can see that AGB stars can be well distinguished from RGB and RSG. The CAGB stars (purple area) are obviously redder than the S-type stars (red area), the OAGB stars are distributed in a relatively large region, and there is no obvious feature on the CMD. The distribution of 606 predicted S-type stars are consistent with the known S-type stars. In addition, the red ``$\times$" denotes 7 stars with $M_{bol}<-7.1$, and we can see that they deviate significantly from most of the S-type stars.

\subsection{Identification of S-type stars with C/O}

Due to the limitation of the wavelength range of LAMOST MRS, it is difficult to measure the abundance of C and O. In this work, we used two methods to estimate C/O, one using the C and O abundance from APOGEE, and the other using (V-K, J-K) photometric indices matching.

\begin{figure}[ht!]
    \centering
	\includegraphics[width=0.5\textwidth]{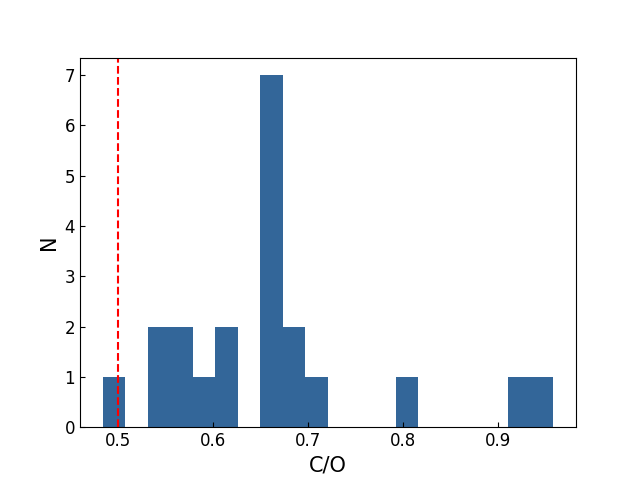} 	
	\caption{The C/O distribution of 21 S-type stars estimated by the [C/Fe] and [O/Fe] of APOGEE DR16, and the C/O=0.5 is denoted by the red dashed line. \label{APOGEE_CO}} 
\end{figure}

APOGEE (\cite{2015AJ....150..148H, 2017AJ....154...94M}) is a median-high resolution (R $\sim$ 22,500) spectroscopic survey in three near-infrared spectral ranges (1.51-1.70 $\mu m$), which allows us to determine the precise atmospheric parameters and chemical abundance, even in highly extinct regions. We cross-matched the 606 S-type stars with APOGEE DR16, got [C/Fe] and [O/Fe] of 21 common stars, and calculated their C/O following the method in \cite{2016ApJ...831...20B}. The C/O distribution of the 21 common stars is shown in Fig.\ref{APOGEE_CO}, there is only one star with $C/O<0.5$, and its C/O is 0.48.

\cite{2017A&A...601A..10V} provided a MARCS S-type star model, which constructed a relationship between V-K and J-K for synthetic spectra with C/O = 0.5, 0.75, 0.90, 0.92, 0.95, 0.97, 0.99, and we used V-K and J-K to estimate the C/Os for the 606 S-type stars. The J and K magnitudes can be obtained through the Point Source Catalog of 2MASS, which containing 470,992,970 sources, and the V magnitude was transformed from magnitudes in g and r band using the method from \cite{2006astro.ph.12034Z}. The g and r magnitudes can come from the Panoramic Survey Telescope and Rapid Response System 1 (Pan-STARRS 1) \citep{chambers2019panstarrs1}, which observes the entire sky north of Dec = -30 deg (the 3$\pi$ survey) in $grizy$. We cross-matched the 606 S-type stars with 2MASS and Pan-STARRS using a radius of 3$''$, and get their J, K, g, and r magnitudes.

\begin{figure}[ht!]
    \centering
	\includegraphics[width=0.5\textwidth]{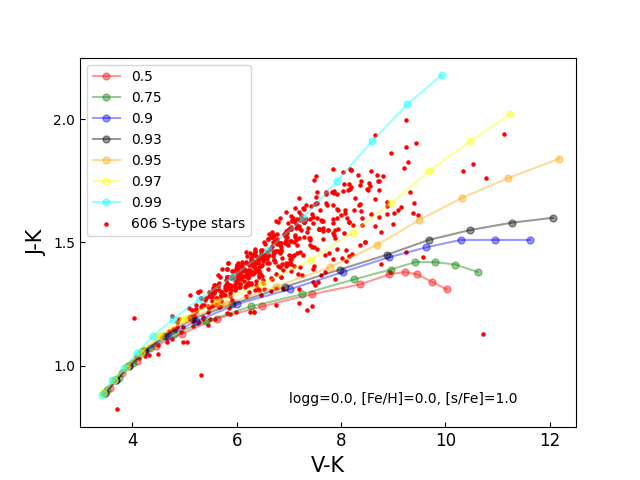}
	\caption{Dereddened V-K and J-K diagram for the 606 S-type stars, as compared to the model colours of different C/O \citep{2017A&A...601A..10V}. \label{Presstar_extinction}}
\end{figure}

\begin{figure}[ht!]
    \centering
	\includegraphics[width=0.5\textwidth]{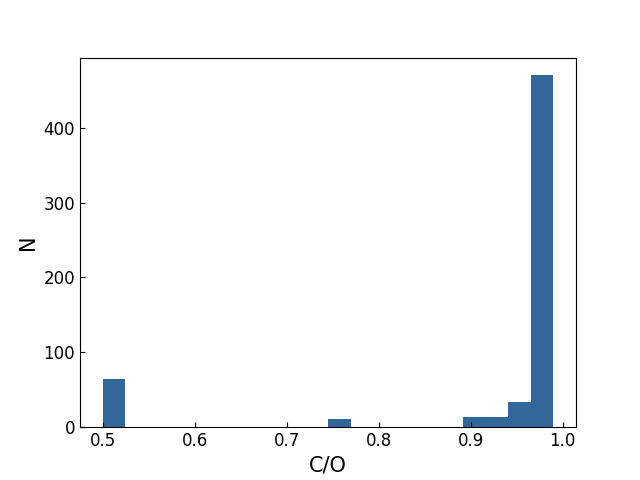}
	\caption{The C/O distribution of 606 S-type stars estimated by the MARCS S-type star model \citep{2017A&A...601A..10V}, which constructed a relationship between V-K and J-K for synthetic spectra with C/O = 0.5, 0.75, 0.90, 0.92, 0.95, 0.97, 0.99. \label{COratio}}
\end{figure}

The extinction of J, K, g, and r magnitudes were estimated by the Python package $dustmaps$ \footnote{https://pypi.org/project/dustmaps}, which needs to the heliocentric distances, right ascension and declination as inputs \citep{2019ApJ...887...93G}, the dereddened J-K and V-K for 606 S-type stars is shown in Fig.\ref{Presstar_extinction}, and the colors of MARCS S-type star models with different C/Os are also shown in Fig.\ref{Presstar_extinction}. Then, the distances between the dereddened colors (J-K and V-K) of 606 S-type stars and  the MARCS models with different C/Os were calculated, and the C/O of the model with the shortest distance was regarded as the C/O of each S-type star. The distribution of 606 S-type stars C/O is showed in Fig.\ref{COratio} and we can see most of the C/O values are larger than 0.9 as expected, which confirms the reliability of the 606 S-type stars to some extent.

\subsection{Cross-match with the SIMBAD}

SIMBAD is the acronym for Set of Identifications, Measurements and Bibliography for Astronomical Data, which is a dynamic database and update every working day. The purpose of Simbad is to provide information for astronomical objects of interest which have been studied in the literature.

The TOPCAT was used to cross-match 606 S-type stars with SIMBAD using a radius of 3 $''$, 380 stars had already been reported in the literatures, and their detailed information is shown in Table \ref{tab:SIMBAD_inf}. The first and second columns of this table list the main type and other types of the 380 stars, and we can see that they were divided into 12 main types. The third column represents the number of each main type, and the extended explanations of each main type is shown in Notes. From the main type column, a fraction of the 380 stars are indeed variables as expected.

In Table \ref{tab:SIMBAD_inf}, there are 64 S-type stars identified in literatures and 3 S-type candidates. After further check, 63 of the 64 stars were included in the eight S-type star catalogs of Section \ref{sample}, the remaining one was classified as an S-type star by \cite{2009MNRAS.400.1413W} based on near-IR spectra, and 3 S candidates were reported in \cite{1975AbaOB..47....3D}. Thus, except for the above 67 S-type stars/candidates, the other 539 were firstly reported as S-type stars in this work.

\begin{table}[ht] 	
	\caption{The information from the SIMBAD for the 380 stars.}\label{tab:SIMBAD_inf} 	\begin{threeparttable}
		\begin{tabular}{lcc} 			\hline \hline 			
			$\mathrm{main\_type}$ & $\mathrm{other\_types}$ & Number  \\ \hline 
			Star \tnote{a}& * $\mid$ IR & 126  \\
			V* \tnote{b} & * $\mid$ IR* & 71 \\
			S* \tnote{c} & LP* $\mid$ V* $\mid$ IR $\mid$ * & 64 \\
			LPV* \tnote{d} & * $\mid$ LP* $\mid$ V* $\mid$ IR & 59 \\ 
			$Candidate.LP*$ \tnote{e} & * $\mid$ IR $\mid$ V* & 41 \\
			Mira \tnote{f} & LP* $\mid$ V* $\mid$ IR $\mid$ * & 10 \\
			$Candidate.S*$ \tnote{g} & S*? $\mid$ IR $\mid$ * $\mid$ S* & 3 \\
			AGB* \tnote{h} & LP* $\mid$ V* $\mid$ IR $\mid$ * & 1 \\
			NIR \tnote{i} & * $\mid$ IR & 2  \\ 
			$Candidate.C*$ \tnote{j} & C*? $\mid$ IR $\mid$ *  & 1 \\
			$Candidate.EB*$ \tnote{k} & * $\mid$ IR* & 1 \\
			$Orion.V*$ \tnote{l} & Or* $\mid$ V* $\mid$ IR & 1  \\ \hline 
		\end{tabular} 	
		\begin{tablenotes} 			
			\footnotesize 	
			\item $^a$ Star
			\item $^b$ Variable star
			\item $^c$ S-type star
			\item $^d$ Long period variable star
			\item $^e$ Long period variable candidate
			\item $^f$ Variable star of Mira Cet type
			\item $^g$ Possible S-type star
			\item $^h$ Asymptotic giant branch star
			\item $^i$ Near infrared source
			\item $^j$ Possible carbon star
			\item $^k$ Eclipsing binary candidate
			\item $^l$ Variable star of Orion type
		\end{tablenotes} 
	\end{threeparttable}
\end{table}

\subsection{The catalog of S-type stars}

After the above further identifications, we constructed a catalog of 24 columns to list parameters for the 606 S-type stars in Table \ref{tab:measured_Inf}. In this table, only 21 rows of the catalog are shown, which are sorted according to the C/O measured by APOGEE, and the full catalog can be available from the link: \href{https://doi.org/10.12149/101097}{https://doi.org/10.12149/101097}. For convenience, we have divided this table into two parts in the paper. The upper part of the table mainly includes the information we calculated, and the bottom part shows the result after cross-matching with Gaia DR2 and 2MASS. The three columns ``TiO", ``ZrO" and ``class" in the table will be discussed in Section \ref{sec:discuss}. The ``-" indicates no measurement.

\begin{splitdeluxetable*}{ccccccccccccBcccccccccccc}
    \tabletypesize{\scriptsize}
    \tablewidth{1pt} 
    \tablecaption{The S-type star catalog constructed in this work. \label{tab:measured_Inf}}
    \tablehead{LAMOST Designation \tnote{a} & ra  & dec  & median snr & TiO & ZrO & class & $C/O_{(A)} $ & $C/O_{(M)}$ & $\rm M_{bol}$  & main\_type  & Sptype & parallax & e\_parallax & G & G$_{BP}$ & G$_{RP}$ & BP-RP  & Jmag  & e\_Jmag & Hmag  & e\_Hmag & Kmag  & e\_Kmag}
    \colnumbers
    \startdata
    J041801.06+603312.0 & 64.5044480  & 60.55333400 & 34  & 0.06  & 0.33 & -    & 0.48 & 0.99 & -1.18 & Star           &    & 0.184 & 0.067 & 13.50 & 15.71 & 12.14 & 3.56 & 9.62  & 0.02 & 8.43 & 0.02 & 8.04 & 0.02\\
    J060547.07+243455.7 & 91.4461640  & 24.58215800 & 143  & 0.13  & 0.27 & E    & 0.53 & 0.99 & -2.21 & Star           &    & 0.278 & 0.080 & 11.51 & 13.36 & 10.23 & 3.13 & 8.04  & 0.02 & 7.02 & 0.02 & 6.64 & 0.02\\
    J025505.05+532412.2 & 43.7710510  & 53.40341600 & 23  & 0.21  & 0.26 & E    & 0.55 & 0.99 & -2.15 & V*             &    & 0.235 & 0.070 & 12.04 & 14.01 & 10.73 & 3.29 & 8.50  & 0.03 & 7.43 & 0.02 & 7.07 & 0.03\\
    J065944.76+041148.4 & 104.9365400 & 4.19678180  & 59  & 0.20  & 0.26 & -    & 0.57 & 0.97 & -1.99 & Star           &    & 0.328 & 0.080 & 11.43 & 13.24 & 10.16 & 3.08 & 8.09  & 0.03 & 7.19 & 0.04 & 6.78 & 0.02\\
    J062437.77+221951.7 & 96.1573890  & 22.33104800 & 32  & 0.21  & 0.28 & -    & 0.58 & 0.97 & -1.67 & V*             &    & 0.190 & 0.175 & 13.01 & 15.64 & 11.57 & 4.07 & 8.85  & 0.03 & 7.62 & 0.05 & 7.22 & 0.02\\
    J004700.39+583122.1 & 11.7516590  & 58.52283000 & 34  & 0.15  & 0.26 & E    & 0.59 & 0.99 & -1.47 & Star           & M3 & 0.377 & 0.076 & 11.58 & 13.33 & 10.34 & 2.99 & 8.27  & 0.03 & 7.37 & 0.06 & 6.99 & 0.02\\
    J064024.65+234358.4 & 100.1027400 & 23.73289400 & 208  & 0.04  & 0.40 & E    & 0.60 & 0.97 & -2.88 & S*             & S & 0.234 & 0.072 & 10.98 & 12.45 & 9.85  & 2.60 & 8.08  & 0.02 & 7.24 & 0.10 & 6.85 & 0.02 \\
    J074420.24+163110.6 & 116.0843400 & 16.51962400 & 94  & 0.12  & 0.30 & $\mathrm{E_L}$ & 0.62 & 0.93 & -4.26 & V*             &   & 0.058 & 0.051 & 12.63 & 13.93 & 11.54 & 2.39 & 9.92  & 0.02 & 9.07 & 0.02 & 8.77 & 0.02 \\
    J045324.01+490003.5 & 73.3500680  & 49.00097300 & 95  & 0.03  & 0.49 & E    & 0.66 & 0.99 & -1.34 & V*             &   & 0.294 & 0.080 & 12.38 & 14.66 & 11.05 & 3.61 & 8.73  & 0.03 & 7.61 & 0.03 & 7.18 & 0.03 \\
    J052710.36+262539.0 & 81.7931860  & 26.42752000 & 32  & 0.03  & 0.35 & -    & 0.66 & 0.99 & -3.23 & V*             &   & 0.078 & 0.080 & 13.35 & 15.24 & 12.06 & 3.17 & 9.80  & 0.02 & 8.65 & 0.03 & 8.28 & 0.03 \\
    J061403.24-054335.7 & 93.5135200  & -5.72658700 & 50  & 0.20  & 0.34 & -    & 0.66 & 0.99 & -5.87 & -              & - & 0.033 & 0.084 & 12.58 & 14.61 & 11.29 & 3.32 & 9.18  & 0.03 & 8.14 & 0.02 & 7.75 & 0.03 \\
    J053357.01+320446.3 & 83.4875780  & 32.07953600 & 26  & 0.04  & 0.34 & -    & 0.66 & 0.99 & -3.82 & -              & - & 0.040 & 0.087 & 14.23 & 16.81 & 12.79 & 4.02 & 10.02 & 0.02 & 8.80 & 0.02 & 8.30 & 0.02 \\
    J042645.72+495356.6 & 66.6905140  & 49.89907400 & 72  & -0.03 & 0.35 & -    & 0.66 & 0.99 & -1.12 & Star           &.  & 0.212 & 0.106 & 13.14 & 15.31 & 11.79 & 3.52 & 9.18  & 0.03 & 7.96 & 0.05 & 7.52 & 0.03\\
    J192453.00+382740.2 & 291.2208500 & 38.46118500 & 52  & 0.21  & 0.25 & -    & 0.66 & 0.75 & -2.24 & LPV*           &   & 0.311 & 0.058 & 11.34 & 13.11 & 10.08 & 3.04 & 8.07  & 0.02 & 7.18 & 0.02 & 6.88 & 0.02 \\
    J041027.17+275343.8 & 62.6132300  & 27.89551000 & 250  & 0.08  & 0.33 & $\mathrm{E_L}$ & 0.67 & 0.97 & -2.99 & V*             &   & 0.159 & 0.047 & 11.95 & 13.48 & 10.77 & 2.71 & 8.98  & 0.02 & 8.09 & 0.06 & 7.75 & 0.03 \\
    J024840.34+522851.1 & 42.1681200  & 52.48087700 & 80  & -0.02  & 0.40 & -    & 0.68 & 0.97 & -1.42 & Star           &.  & 0.222 & 0.042 & 12.55 & 14.00 & 11.41 & 2.58 & 9.61  & 0.02 & 8.69 & 0.03 & 8.39 & 0.02 \\
    J063403.50+230152.3 & 98.5145980  & 23.03119700 & 164 & 0.10  & 0.34 & -    & 0.68 & 0.9  & -2.21 & Star           &   & 0.338 & 0.088 & 10.98 & 12.49 & 9.83  & 2.67 & 8.00  & 0.03 & 7.09 & 0.02 & 6.80 & 0.02 \\
    J044915.36+444614.8 & 72.3140290  & 44.77078600 & 141 & -0.02 & 0.32 & E    & 0.70 & 0.99 & -1.07 & Star           &   & 0.342 & 0.059 & 12.06 & 13.71 & 10.85 & 2.87 & 8.74  & 0.03 & 7.69 & 0.03 & 7.33 & 0.02 \\
    J042243.70+172919.7 & 65.6821190  & 17.48880900 & 95  & 0.11  & 0.38 & E    & 0.81 & 0.97 & -2.12 & Candidate\_LP* &   & 0.204 & 0.100 & 12.38 & 14.44 & 11.08 & 3.36 & 8.89  & 0.02 & 7.96 & 0.02 & 7.55 & 0.02 \\
    J054344.26+143843.3 & 85.9344190  & 14.64536600 & 49  & -0.02  & 0.44 & -    & 0.92 & 0.99 & -2.36 & Star           &   & 0.213 & 0.060 & 11.95 & 13.68 & 10.74 & 2.94 & 8.74  & 0.02 & 7.74 & 0.02 & 7.38 & 0.02 \\
    J061207.47-063700.2 & 93.0311400  & -6.61672300 & 80  & 0.18  & 0.32 & $\mathrm{E_L}$ & 0.96 & 0.99 & -3.66 & V*             &   & 0.117 & 0.076 & 12.04 & 14.14 & 10.72 & 3.42 & 8.50  & 0.02 & 7.43 & 0.03 & 7.07 & 0.02 \\
    \enddata
 	\tablecomments{\\
 	    Column (1) The designation from LAMOST DR9.\\
 	    Column (4) The median value of all pixel single-to-noises (S/Ns) in the single exposure red band spectrum, which is the highest S/N of multiple epoch for one star.\\
 	    Column (5) and (6) The band indices of TiO and ZrO obtained from the red band spectrum with the highest S/N, which will be discussed in Section \ref{sec:discuss}.\\
 	    Column (7) The class of binary candidates, which will be discussed in Section \ref{sec:discuss}.\\
 	    Column (8) The C/O estimated by the [C/Fe] and [O/Fe] from APOGEE.\\
 	    Column (9) The C/O estimated by the MARCS S-type star model.\\
 	    Column (10) The absolute bolometric magnitude.\\
 	    Column (11) and (12) The main type and spectral type from the SIMBAD.\\
 	    Column (13) - (18) The parallax and uncertainty, the three broad-band magnitude, and the BP - RP color from Gaia DR2.\\
 	    Column (19) - (24) The J, H, and K magnitudes and their uncertainties from 2MASS.}
\end{splitdeluxetable*}

\begin{figure}[ht!] 
	\centering
        \includegraphics[width=0.5\textwidth]{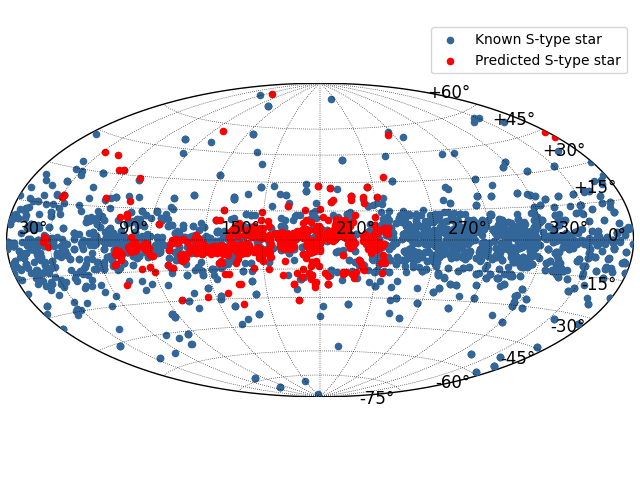}
	\caption{The spatial distribution of the 606 S-type stars in the Galactic coordinates. \label{spatial_distribution}} 
\end{figure}

The spatial distribution of the 606 S-type stars in Galactic coordinates is plotted in Fig.\ref{spatial_distribution}, and the red and blue dots represent predicted and known S-type stars, respectively. In the direction of galactic latitude, the distribution of the two samples of stars is relatively consistent, and they are both within $-15^{\circ} < b < +15^{\circ}$. In the galactic longitude direction, the known S-type stars are distributed from $0^{\circ}$ to $360^{\circ}$, and the S-type stars in this work are mainly focus on the range of $60^{\circ} < l < 210^{\circ}$.

\section{Discussion}\label{sec:discuss}

\subsection{Analyzing the effect of TiO and atmospheric parameters on ZrO}

\begin{figure*}[ht]
	\centering
    \includegraphics[width=1.0\textwidth]{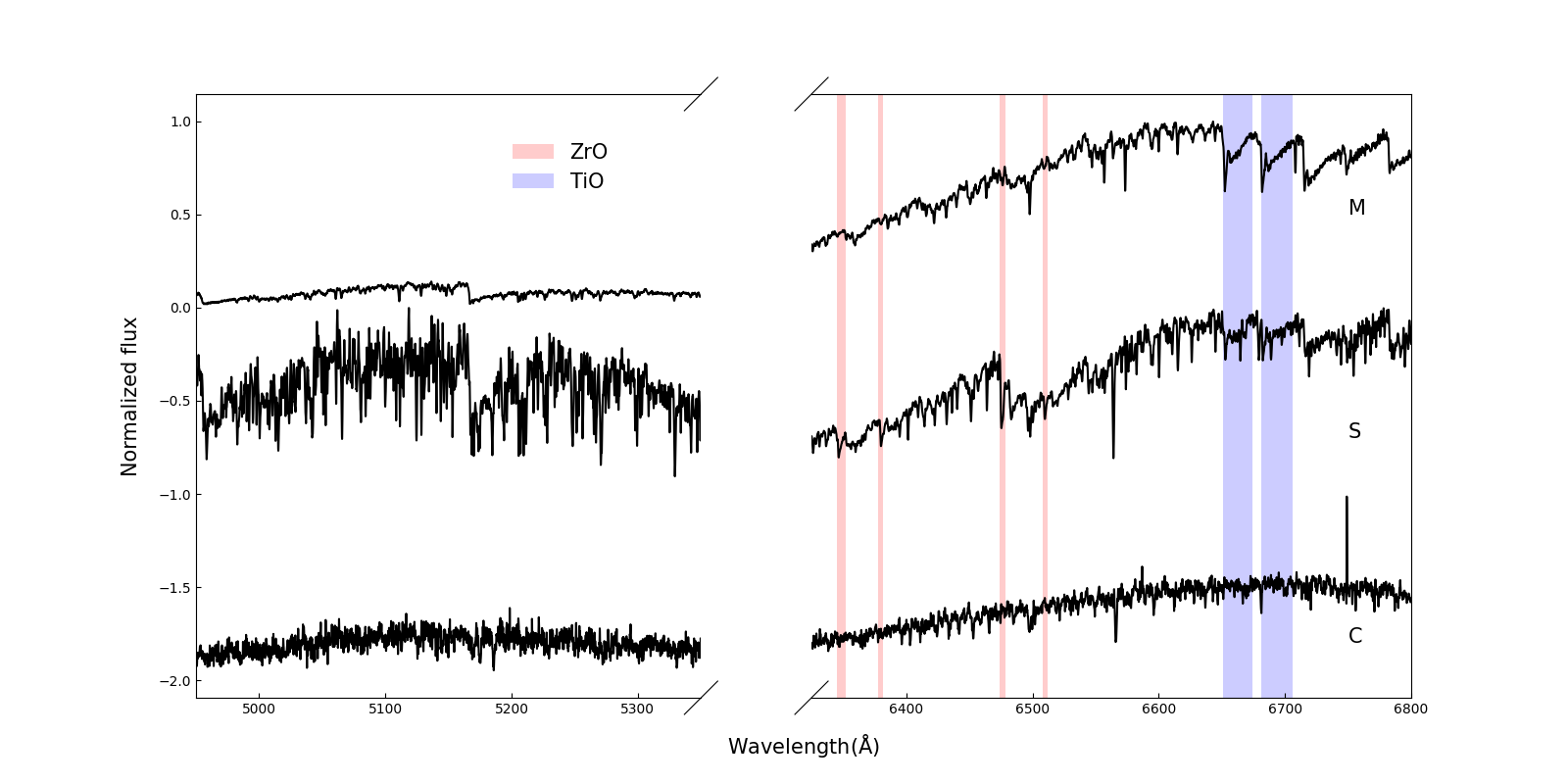}
	\caption{From top to bottom, the spectra of the M-S-C evolution sequence are represented. The M-type stars have strong TiO bands, the S-type stars has strong ZrO and TiO relatively weak, and both ZrO and TiO bands are disappeared in the carbon star. The red and blue areas are ZrO and TiO bands, respectively. \label{Sstar_template}}
\end{figure*}

Figure \ref{Sstar_template} shows the example spectra of an M-type, an S-type and a Carbon star from top to bottom, and the left and right parts are their blue and red band spectra, respectively. It can be seen from the figure that there is no obvious spectral features at the blue band spectra, both ZrO (red area) and TiO (purple area) are obvious in the red band spectra, and the ZrOs of S-type stars are much stronger.

\begin{figure}[ht]
	\centering
    \includegraphics[width=0.5\textwidth]{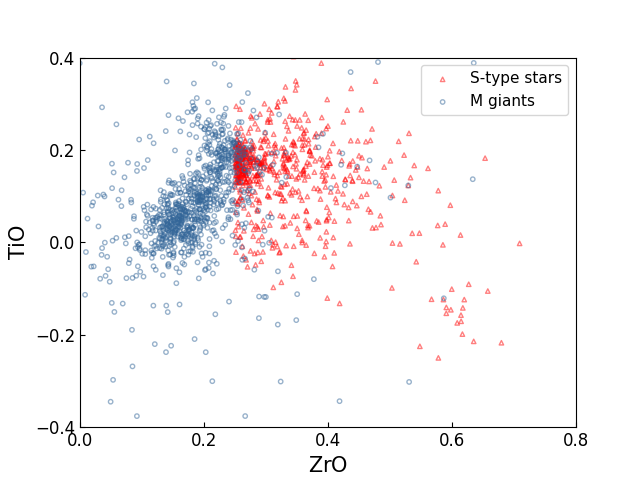}
	\caption{Measured $\rm B_{ZrO}$ and $\rm B_{TiO}$ band indices of 606 S-type stars (red) and 1000 M giants (blue), respectively. \label{ZrO_VS_TiO}}
\end{figure}

The TiO band indices of 606 S-type stars were calculated in the same way as ZrO, which are also included in our S-type star catalog of Table \ref{tab:measured_Inf}, and the band and continuum windows are listed in Table \ref{tab:TiOband}. For Comparison, the TiO and ZrO band indices of 1000 M giants from S. Li et al. (in preparation) were also estimated, and the two indices of the 606 S-type stars and 1000 M giants are displayed in Figure \ref{ZrO_VS_TiO}. It should be note that the TiO and ZrO band indices of each S-type star and M giant were calculated by the red band spectrum with the highest S/N. From this figure, the TiO indices of S-type stars and M giants are overlapped in the range of [-0.4, 0.4], and the ZrOs of S-type stars are obviously stronger than those of M-giant stars. The TiO and ZrO band indices of M giants are mainly distributed in the range of [0.0, 0.25] and [0.1, 0.3], respectively, and there is an obviously positive correlation between ZrO and TiO for M giants. The two band indices for S-type stars mainly focus on the range of [0.0, 0.3] and [0.25, 0.5], there are still a few S-type stars located near TiO = -0.2 and ZrO = 0.6, and there may be a negative correlation between ZrO and TiO for S-type stars, which needs more S-type star samples to verify.

\begin{table}[ht]
	\centering
	\caption{Boundaries of the continuum and band windows of TiO used in the computation of the band indices.}\label{tab:TiOband}
	\begin{threeparttable}
	    \begin{tabular}{ccccc} \hline \hline
			band & $\lambda_{B,i}$ & $\lambda_{B,f}$ & $\lambda_{C,i}$ & $\lambda_{C,f}$ \\ \hline
			$\mathrm{TiO_{1}}$ & 6651.0 & 6674.0 & 6646.0 & 6650.0 \\ 
			$\mathrm{TiO_{2}}$ & 6681.0 & 6706.0 & 6646.0 & 6650.0 \\ \hline
		\end{tabular}
	\end{threeparttable}
\end{table}

\begin{figure}[ht]
	\centering
    \includegraphics[width=0.5\textwidth]{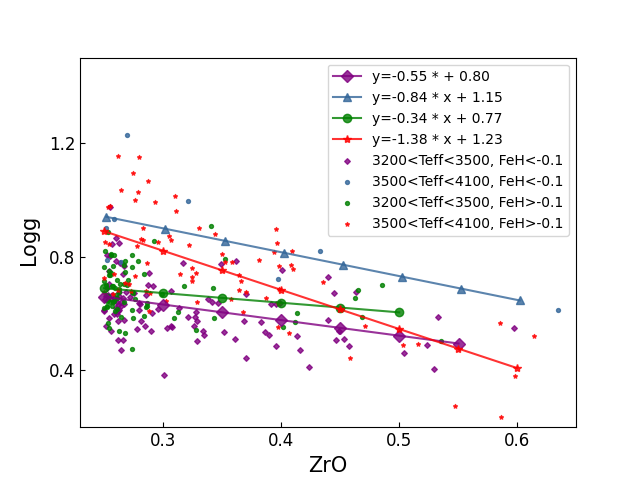}
	\caption{The surface gravity vs. ZrO for the 288 S-type stars with atmospheric parameters given by LAMOST. The points and lines with different colors represent four groups of the S-type stars, and their fitting results with the first order polynomials, respectively. \label{logg_VS_ZrO}}
\end{figure}

The `` LAMOST DR9 Parameter Catalog'' provides the atmospheric parameters of $\rm{T_{eff}}$, log $g$ and [Fe/H] for a fraction of LAMOST MRS spectra, the 606 S-type stars were cross-matched with the catalog, and $\rm{T_{eff}}$, log $g$ and [Fe/H] of 288 stars were obtained. The effect of $\rm{T_{eff}}$, log $g$ and [Fe/H] on ZrO of the 288 S-type stars were analyzed, and we found that log $g$ seems to have the greatest effect on ZrO. The ${T_{\rm eff}}$ of the 288 S-type stars were divided into two bins of [3200 K, 3500 K] and [3500 K, 4100 K], their [Fe/H] were divided into two ranges of $> -0.1$ and $< 0.1$, and the 288 stars were clustered into four groups according to different ${T_{\rm eff}}$ and [Fe/H] ranges. log $g$ and ZrO indices for stars of each group were fitted with the first order polynomials, which are shown by different color lines in Fig.\ref{logg_VS_ZrO}, and the four fitting functions are displayed in the legend. As can be seen from this figure, log $g$ has a negative correlation with ZrO for the four group stars, the slopes of the fitting polynomials are much steeper for the two groups of warmer stars, and ${T_{\rm eff}}$ and [Fe/H] have little effects on ZrO. In future, more S-type stars with accurate atmospheric parameters are needed to further investigate the effects of atmospheric parameters on ZrO.

\subsection{Binary Candidates}

The ``LAMOST MRS general catalog" provides RVs of blue and red band spectra, and the RVs were calibrated by the radial velocity standard stars provided in \cite{2018AJ....156...90H}. \cite{2021PASP..133d4502C} compared the RVs of 9110 objects in both LAMOST and SDSS14/APOGEE, and the dispersion is only around one $\mathrm{km ~s^{-1}}$. In addition, \cite{2021ApJS..256...14Z} measured RVs of 3.8 million single-exposure spectra (about 0.6 million stars) of LAMOST MRS with S/N ratios greater than 5, and propose a robust method to determine the RV zero-point, which has average accuracy of 0.38 $\rm km~ s^{-1}$. Since the co-added spectra of LAMOST MRS are the results of manual processing, which may introduce inevitable problems, so we only use the RVs of single exposure spectra for analysis. Due to the spectrum characteristics of S-type stars are mainly in red band, the RVs after zero-point calibrations are all provided in ``LAMOST MRS general catalog" and \cite{2021ApJS..256...14Z}, and only the calibrated RVs of red band spectra  ($\mathrm{RV\_red_{LAMOST}}$ and $\mathrm{RV\_red_{Zhang}}$) were used.

\begin{figure}[ht]
    \centering
	\includegraphics[width=0.5\textwidth]{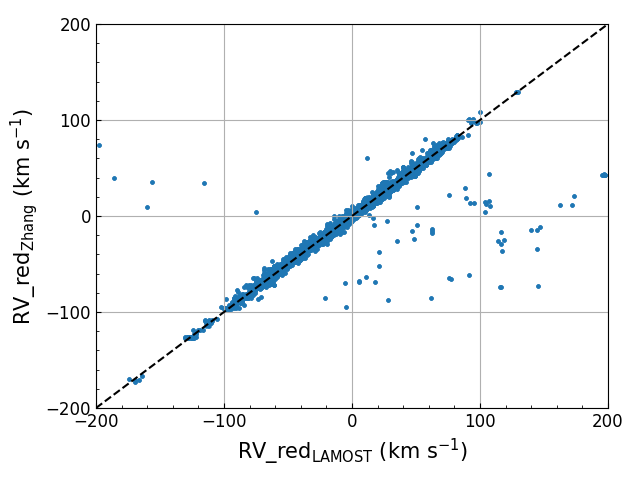}
	\includegraphics[width=0.5\textwidth]{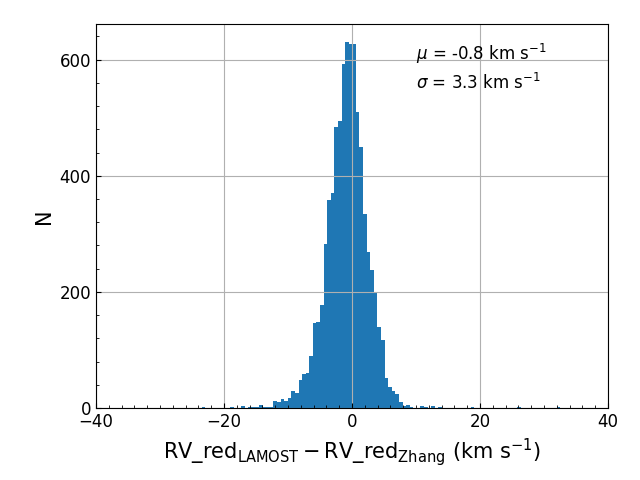}
	\caption{The RV comparison of red band spectra from ``LAMOST MRS general catalog" and the catalog in \cite{2021ApJS..256...14Z} ($\mathrm{RV\_red_{LAMOST}}$ and $\mathrm{RV\_red_{Zhang}}$). The top panel shows the point-to-point comparisons of $\mathrm{RV\_red_{LAMOST}}$ and $\mathrm{RV\_red_{Zhang}}$, and the distribution of their difference is shown in the bottom panel. \label{RV_VS_Zhang}}
\end{figure}

We crossed the 606 S-type stars with ``LAMOST MRS general catalog" and the catalog of \cite{2021ApJS..256...14Z}, get the $\mathrm{RV\_red_{LAMOST}}$ and $\mathrm{RV\_red_{Zhang}}$ of 8728 single-exposure spectra of 506 stars, and the spectra with RV=-9999 $\rm km~ s^{-1}$ were excluded. The comparison of $\mathrm{RV\_red_{LAMOST}}$ and $\mathrm{RV\_red_{Zhang}}$ for 506 stars is shown in the top panel of Figure \ref{RV_VS_Zhang}, and they are relatively consistent. The bottom panel shows the difference distribution of $\mathrm{RV\_red_{LAMOST}}$ and $\mathrm{RV\_red_{Zhang}}$, and the offset and dispersion of the difference are -0.8 $\mathrm{km ~s^{-1}}$ and 3.3 $\mathrm{km ~s^{-1}}$, respectively, when $\mid \mathrm{RV\_red_{LAMOST} - RV\_red_{Zhang} \mid < 40 ~km~ s^{-1}}$.

\begin{figure*}[ht]
    \centering
	\includegraphics[width=0.4\textwidth]{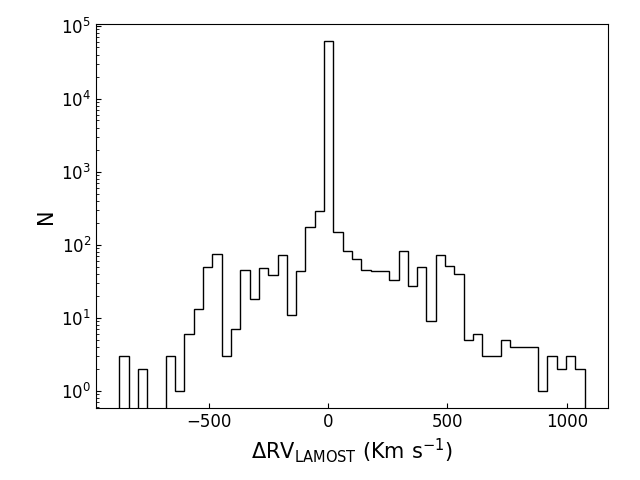}
	\includegraphics[width=0.4\textwidth]{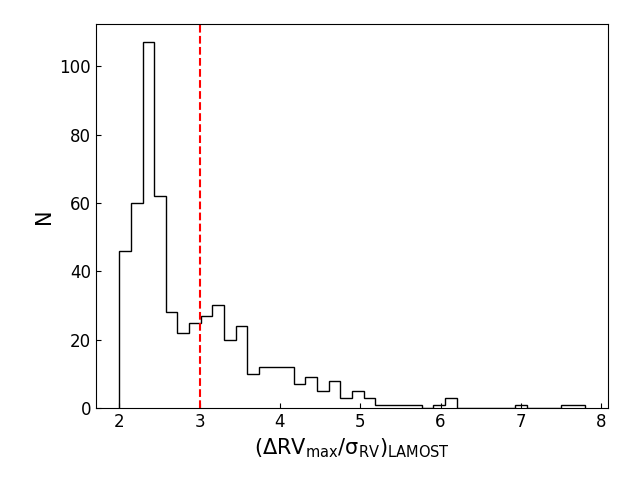}
	\includegraphics[width=0.4\textwidth]{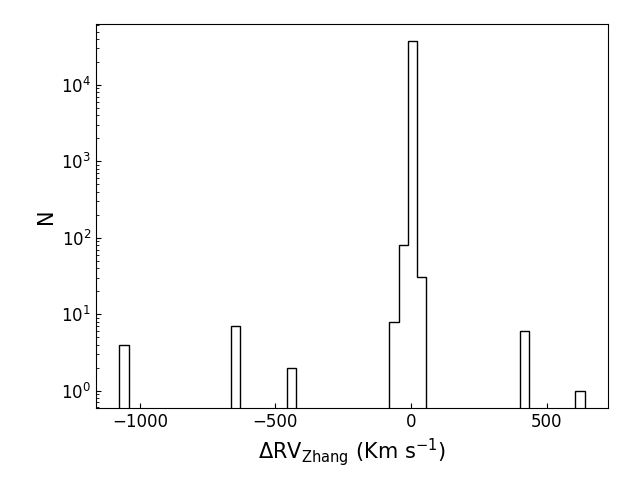}
	\includegraphics[width=0.4\textwidth]{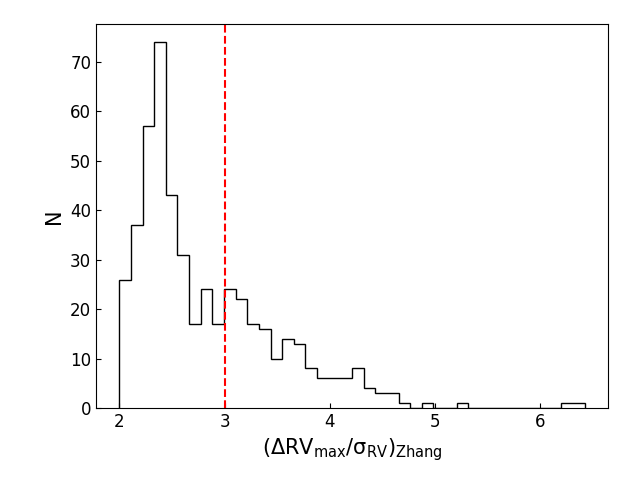}
	\caption{The top and bottom panels give the distributions of $\rm \Delta RV$ and $\rm \Delta RV_{max} / \sigma_{RV}$ estimated by $\mathrm{RV\_red_{LAMOST}}$ and $\mathrm{RV\_red_{Zhang}}$ (introduced in Fig.\ref{RV_VS_Zhang}), respectively, where $\rm \Delta RV$ is the RV difference between any two epochs for an object, $\rm \Delta RV_{max}$ is the maximum of $\rm \Delta RV$, and $\rm \sigma_{RV}$ is the standard deviation of multiple epochs. The red dashed line denotes the criterion of $\rm \Delta RV_{max} = 3.0 ~ \sigma_{RV}$ to select binary candidates. \label{deltaRV_VS_sigmaRV}}
\end{figure*}

\cite{2020ApJS..249...22T} used $\rm \Delta RV_{max} > 3.0 ~ \sigma_{RV}$ to identify binary stars, where $\rm \Delta RV$ is the RV difference between any two epochs for an object, $\rm \Delta RV_{max}$ is the maximum of $\rm \Delta RV$, and $\rm \sigma_{RV}$ is the standard deviation of multiple epochs. We used $\mathrm{RV\_red_{LAMOST}}$ and $\mathrm{RV\_red_{Zhang}}$ to calculate the $\rm \Delta RV$ and $\rm \Delta RV_{max} / \sigma_{RV}$ for 506 S-type stars, which are shown in Figure \ref{deltaRV_VS_sigmaRV}. According to the criterion of $\rm \Delta RV_{max} > 3.0 ~ \sigma_{RV}$, there are a total of 238 binary candidates found by $\mathrm{RV\_red_{LAMOST}}$ and $\mathrm{RV\_red_{Zhang}}$, and they are marked in the ``class" column of Table \ref{tab:measured_Inf} using  ``E", ``E$_{\rm L}$" and ``E$_{\rm Z}$", which denotes that they were classified as extrinsic S-type (binary) candidates using both $\mathrm{RV\_red_{LAMOST}}$ and $\mathrm{RV\_red_{Zhang}}$, only $\mathrm{RV\_red_{LAMOST}}$, or only $\mathrm{RV\_red_{Zhang}}$, respectively. Among the 238 binary candidates, there are 127, 73, and 38 stars classified as ``E", ``E$_{\rm L}$" and ``E$_{\rm Z}$", respectively, and the 127 stars with class ``E" are more likely to be binary candidates because they satisfy the criterion of \cite{2020ApJS..249...22T} no matter $\mathrm{RV\_red_{LAMOST}}$ or $\mathrm{RV\_red_{Zhang}}$ was used. ``-" in the ``class" column represents the other stars that were not considered as binary candidates in this work, due to the relatively few observations or the relatively low accuracies of RV measurements.

\section{Summary} \label{sec:summary}

In this paper, we preliminarily selected S-type star candidates from the low-temperature giant sample of LAMOST DR9 with the criterion of ZrO band indices, further used the iterative Support Vector Machine (SVM) algorithm to select S-type stars from the candidates, and 613 S-type stars were finally selected after the iterative SVM. We investigate these stars, and the main conclusions are:


\begin{enumerate}
\item [(1)] In general, the absolute bolometric magnitude of S-type stars is larger than -7.1 \citep{1983ApJ...272...99W}, we calculated the $\rm M_{bol}$ for 613 S-type stars, and found there were 7 stars with $\rm M_{bol} < -7.1$. The remaining 606 stars were our final S-type samples, 67 S-type stars/candidates among them were already identified in literature through cross-matching with SIMBAD, and the other 539 stars were firstly reported in this paper. In the color-magnitude diagram of (J-K) and $\rm M_{K}$, the 606 S-type stars are distributed in the AGB stage, and are also consistent with the distribution of known S-type stars.
\item [(2)] We used two methods to evaluate C/O of 606 S-type stars. Firstly, the C/Os of 21 S-type stars were obtained by the C and O abundances from APOGEE DR16, and 20 of them have C/O $>$ 0.5. Secondly, a MARCS model of S-type stars is provided by \cite{2017A&A...601A..10V}, which can be used to estimate C/O (from 0.5 to 0.99) with V-K and J-K, the C/Os of 606 S-type stars were estimated with this model, and most of the C/O values are larger than 0.9.
\item [(3)] The band indices of TiO and ZrO for 606 S-type stars and 1000 M giants were calculated, there may be a negative correlation between ZrO and TiO for S-type stars, which is different from the obvious positive correlation of the two bands for M giants. In addition, the effects of atmospheric parameters on ZrO were analyzed, log g has a negative correlation with ZrO, which has the greatest effect on ZrO compared with Teff and [Fe/H], and the slopes of the negative correlation are much steeper for warmer stars. More S-type stars are needed in future to verify the relationship of TiO vs. ZrO, and log g vs. ZrO.
\item [(4)] We cross-matched 606 S-type stars with ``the LAMOST MRS general catalog" and the catalog of \cite{2021ApJS..256...14Z}, got 506 common stars with zero point calibrated RVs of red band spectra ($\mathrm{RV\_red_{LAMOST}}$ and $\mathrm{RV\_red_{Zhang}}$). According to the criterion of \cite{2020ApJS..249...22T}, there are a total of 238 binary candidates found by $\mathrm{RV\_red_{LAMOST}}$ and $\mathrm{RV\_red_{Zhang}}$.
\end{enumerate}

We constructed a catalog for 606 S-type stars, which includes 24 columns such as the band indices of TiO and ZrO, the C/Os obtained by two methods, the cross-matched results with SIMBAD, Gaia, and 2MASS, and it is available from the link: \href{https://doi.org/10.12149/101097}{https://doi.org/10.12149/101097}.



\section{Acknowledgements}
We thank the anonymous referee for his/her valuable comments and suggestions. We also thank Guo Yan-Xin and Kong Xiao for helpful discussions.
This work is supported by National Science Foundation of China (Nos U 1931209, 12003050) and National Key R\&D Program of China (No. 2019YFA 0405502), China Manned Space Project ( Nos CMS-CSST-2021-A10, CMS-CSST-2021-B05). Guoshoujing Telescope (the Large Sky Area Multi-Object Fiber Spectroscopic Telescope, LAMOST) is a National Major Scientific Project built by the Chinese Academy of Sciences. Funding for the Project has been provided by the National Development and Reform Commission. LAMOST is operated and managed by the National Astronomical Observatories, Chinese Academy of Sciences.
This research makes use of data from the European Space Agency (ESA) mission Gaia, processed by the Gaia Data Processing and Anal- ysis Consortium. This research also makes use of Astropy, a community-developed core Python package for Astronomy \citep{2013A&A...558A..33A}, the TOPCAT tool \citep{2005ASPC..347...29T} and the VizieR catalog access tool and the Simbad database, operated at CDS, Strasbourg, France.


\bibliography{reference}{}

\begin{thebibliography}{}
\expandafter\ifx\csname natexlab\endcsname\relax\def\natexlab#1{#1}\fi
\providecommand{\url}[1]{\href{#1}{#1}}
\providecommand{\dodoi}[1]{doi:~\href{http://doi.org/#1}{\nolinkurl{#1}}}
\providecommand{\doeprint}[1]{\href{http://ascl.net/#1}{\nolinkurl{http://ascl.net/#1}}}
\providecommand{\doarXiv}[1]{\href{https://arxiv.org/abs/#1}{\nolinkurl{https://arxiv.org/abs/#1}}}

\bibitem[{{Ake}(1979)}]{1979ApJ...234..538A}
{Ake}, T.~B. 1979, \apj, 234, 538, \dodoi{10.1086/157527}

\bibitem[{{Astropy Collaboration} {et~al.}(2013){Astropy Collaboration},
  {Robitaille}, {Tollerud}, {Greenfield}, {Droettboom}, {Bray}, {Aldcroft},
  {Davis}, {Ginsburg}, {Price-Whelan}, {Kerzendorf}, {Conley}, {Crighton},
  {Barbary}, {Muna}, {Ferguson}, {Grollier}, {Parikh}, {Nair}, {Unther},
  {Deil}, {Woillez}, {Conseil}, {Kramer}, {Turner}, {Singer}, {Fox}, {Weaver},
  {Zabalza}, {Edwards}, {Azalee Bostroem}, {Burke}, {Casey}, {Crawford},
  {Dencheva}, {Ely}, {Jenness}, {Labrie}, {Lim}, {Pierfederici}, {Pontzen},
  {Ptak}, {Refsdal}, {Servillat}, \& {Streicher}}]{2013A&A...558A..33A}
{Astropy Collaboration}, {Robitaille}, T.~P., {Tollerud}, E.~J., {et~al.} 2013,
  \aap, 558, A33, \dodoi{10.1051/0004-6361/201322068}

\bibitem[{{Blanco} \& {Nassau}(1957)}]{1957ApJ...125..408B}
{Blanco}, V.~M., \& {Nassau}, J.~J. 1957, \apj, 125, 408,
  \dodoi{10.1086/146316}

\bibitem[{{Boyer} {et~al.}(2011){Boyer}, {Srinivasan}, {van Loon}, {McDonald},
  {Meixner}, {Zaritsky}, {Gordon}, {Kemper}, {Babler}, {Block}, {Bracker},
  {Engelbracht}, {Hora}, {Indebetouw}, {Meade}, {Misselt}, {Robitaille},
  {Sewi{\l}o}, {Shiao}, \& {Whitney}}]{2011AJ....142..103B}
{Boyer}, M.~L., {Srinivasan}, S., {van Loon}, J.~T., {et~al.} 2011, \aj, 142,
  103, \dodoi{10.1088/0004-6256/142/4/103}

\bibitem[{{Brewer} \& {Fischer}(2016)}]{2016ApJ...831...20B}
{Brewer}, J.~M., \& {Fischer}, D.~A. 2016, \apj, 831, 20,
  \dodoi{10.3847/0004-637X/831/1/20}

\bibitem[{Chambers {et~al.}(2019)Chambers, Magnier, Metcalfe, Flewelling,
  Huber, Waters, Denneau, Draper, Farrow, Finkbeiner, Holmberg, Koppenhoefer,
  Price, Rest, Saglia, Schlafly, Smartt, Sweeney, Wainscoat, Burgett, Chastel,
  Grav, Heasley, Hodapp, Jedicke, Kaiser, Kudritzki, Luppino, Lupton, Monet,
  Morgan, Onaka, Shiao, Stubbs, Tonry, White, Bañados, Bell, Bender, Bernard,
  Boegner, Boffi, Botticella, Calamida, Casertano, Chen, Chen, Cole, Deacon,
  Frenk, Fitzsimmons, Gezari, Gibbs, Goessl, Goggia, Gourgue, Goldman, Grant,
  Grebel, Hambly, Hasinger, Heavens, Heckman, Henderson, Henning, Holman, Hopp,
  Ip, Isani, Jackson, Keyes, Koekemoer, Kotak, Le, Liska, Long, Lucey, Liu,
  Martin, Masci, McLean, Mindel, Misra, Morganson, Murphy, Obaika, Narayan,
  Nieto-Santisteban, Norberg, Peacock, Pier, Postman, Primak, Rae, Rai, Riess,
  Riffeser, Rix, Röser, Russel, Rutz, Schilbach, Schultz, Scolnic, Strolger,
  Szalay, Seitz, Small, Smith, Soderblom, Taylor, Thomson, Taylor, Thakar,
  Thiel, Thilker, Unger, Urata, Valenti, Wagner, Walder, Walter, Watters,
  Werner, Wood-Vasey, \& Wyse}]{chambers2019panstarrs1}
Chambers, K.~C., Magnier, E.~A., Metcalfe, N., {et~al.} 2019.
\newblock \doarXiv{1612.05560}

\bibitem[{{Chen} {et~al.}(2019){Chen}, {Liu}, \& {Shan}}]{2019AJ....158...22C}
{Chen}, P.~S., {Liu}, J.~Y., \& {Shan}, H.~G. 2019, \aj, 158, 22,
  \dodoi{10.3847/1538-3881/ab2334}

\bibitem[{{Chen} {et~al.}(2021){Chen}, {Luo}, {Chen}, {Du}, {Wang}, {Zuo},
  {Zhang}, {Li}, {Lu}, {Chen}, \& {Qu}}]{2021PASP..133d4502C}
{Chen}, X.-L., {Luo}, A.~L., {Chen}, J.-J., {et~al.} 2021, \pasp, 133, 044502,
  \dodoi{10.1088/1538-3873/abe0ac}

\bibitem[{{Cortes} \& {Vapnik}(1995)}]{1995ML...20....273}
{Cortes}, C., \& {Vapnik}, V. 1995, Machine Learning, 20, 273,
  \dodoi{10.1007/BF00994018}

\bibitem[{{Cui} {et~al.}(2012){Cui}, {Zhao}, {Chu}, {Li}, {Li}, {Zhang}, {Su},
  {Yao}, {Wang}, {Xing}, {Li}, {Zhu}, {Wang}, {Gu}, {Luo}, {Xu}, {Zhang},
  {Liu}, {Zhang}, {Yang}, {Cao}, {Chen}, {Chen}, {Chen}, {Chen}, {Chu}, {Feng},
  {Gong}, {Hou}, {Hu}, {Hu}, {Hu}, {Jia}, {Jiang}, {Jiang}, {Jiang}, {Jin},
  {Li}, {Li}, {Li}, {Liu}, {Liu}, {Lu}, {Mao}, {Men}, {Qi}, {Qi}, {Shi},
  {Tang}, {Tao}, {Wang}, {Wang}, {Wang}, {Wang}, {Wang}, {Wang}, {Wang},
  {Wang}, {Wang}, {Wang}, {Wang}, {Wang}, {Xu}, {Xu}, {Yang}, {Yu}, {Yuan},
  {Yuan}, {Zhai}, {Zhang}, {Zhang}, {Zhang}, {Zhao}, {Zhou}, {Zhou}, {Zhu}, \&
  {Zou}}]{2012RAA....12.1197C}
{Cui}, X.-Q., {Zhao}, Y.-H., {Chu}, Y.-Q., {et~al.} 2012, Research in Astronomy
  and Astrophysics, 12, 1197, \dodoi{10.1088/1674-4527/12/9/003}

\bibitem[{{Dolidze}(1975)}]{1975AbaOB..47....3D}
{Dolidze}, M.~V. 1975, Abastumanskaia Astrofizicheskaia Observatoriia
  Byulleten, 47, 3

\bibitem[{{Gaia Collaboration} {et~al.}(2016){Gaia Collaboration}, {Prusti},
  {de Bruijne}, {Brown}, {Vallenari}, {Babusiaux}, {Bailer-Jones}, {Bastian},
  {Biermann}, {Evans}, {Eyer}, {Jansen}, {Jordi}, {Klioner}, {Lammers},
  {Lindegren}, {Luri}, {Mignard}, {Milligan}, {Panem}, {Poinsignon},
  {Pourbaix}, {Randich}, {Sarri}, {Sartoretti}, {Siddiqui}, {Soubiran},
  {Valette}, {van Leeuwen}, {Walton}, {Aerts}, {Arenou}, {Cropper}, {Drimmel},
  {H{\o}g}, {Katz}, {Lattanzi}, {O'Mullane}, {Grebel}, {Holland}, {Huc},
  {Passot}, {Bramante}, {Cacciari}, {Casta{\~n}eda}, {Chaoul}, {Cheek}, {De
  Angeli}, {Fabricius}, {Guerra}, {Hern{\'a}ndez}, {Jean-Antoine-Piccolo},
  {Masana}, {Messineo}, {Mowlavi}, {Nienartowicz}, {Ord{\'o}{\~n}ez-Blanco},
  {Panuzzo}, {Portell}, {Richards}, {Riello}, {Seabroke}, {Tanga},
  {Th{\'e}venin}, {Torra}, {Els}, {Gracia-Abril}, {Comoretto},
  {Garcia-Reinaldos}, {Lock}, {Mercier}, {Altmann}, {Andrae}, {Astraatmadja},
  {Bellas-Velidis}, {Benson}, {Berthier}, {Blomme}, {Busso}, {Carry},
  {Cellino}, {Clementini}, {Cowell}, {Creevey}, {Cuypers}, {Davidson}, {De
  Ridder}, {de Torres}, {Delchambre}, {Dell'Oro}, {Ducourant}, {Fr{\'e}mat},
  {Garc{\'\i}a-Torres}, {Gosset}, {Halbwachs}, {Hambly}, {Harrison}, {Hauser},
  {Hestroffer}, {Hodgkin}, {Huckle}, {Hutton}, {Jasniewicz}, {Jordan},
  {Kontizas}, {Korn}, {Lanzafame}, {Manteiga}, {Moitinho}, {Muinonen},
  {Osinde}, {Pancino}, {Pauwels}, {Petit}, {Recio-Blanco}, {Robin}, {Sarro},
  {Siopis}, {Smith}, {Smith}, {Sozzetti}, {Thuillot}, {van Reeven}, {Viala},
  {Abbas}, {Abreu Aramburu}, {Accart}, {Aguado}, {Allan}, {Allasia},
  {Altavilla}, {{\'A}lvarez}, {Alves}, {Anderson}, {Andrei}, {Anglada Varela},
  {Antiche}, {Antoja}, {Ant{\'o}n}, {Arcay}, {Atzei}, {Ayache}, {Bach},
  {Baker}, {Balaguer-N{\'u}{\~n}ez}, {Barache}, {Barata}, {Barbier}, {Barblan},
  {Baroni}, {Barrado y Navascu{\'e}s}, {Barros}, {Barstow}, {Becciani},
  {Bellazzini}, {Bellei}, {Bello Garc{\'\i}a}, {Belokurov}, {Bendjoya},
  {Berihuete}, {Bianchi}, {Bienaym{\'e}}, {Billebaud}, {Blagorodnova},
  {Blanco-Cuaresma}, {Boch}, {Bombrun}, {Borrachero}, {Bouquillon}, {Bourda},
  {Bouy}, {Bragaglia}, {Breddels}, {Brouillet}, {Br{\"u}semeister},
  {Bucciarelli}, {Budnik}, {Burgess}, {Burgon}, {Burlacu}, {Busonero}, {Buzzi},
  {Caffau}, {Cambras}, {Campbell}, {Cancelliere}, {Cantat-Gaudin}, {Carlucci},
  {Carrasco}, {Castellani}, {Charlot}, {Charnas}, {Charvet}, {Chassat},
  {Chiavassa}, {Clotet}, {Cocozza}, {Collins}, {Collins}, {Costigan}, {Crifo},
  {Cross}, {Crosta}, {Crowley}, {Dafonte}, {Damerdji}, {Dapergolas}, {David},
  {David}, {De Cat}, {de Felice}, {de Laverny}, {De Luise}, {De March}, {de
  Martino}, {de Souza}, {Debosscher}, {del Pozo}, {Delbo}, {Delgado},
  {Delgado}, {di Marco}, {Di Matteo}, {Diakite}, {Distefano}, {Dolding}, {Dos
  Anjos}, {Drazinos}, {Dur{\'a}n}, {Dzigan}, {Ecale}, {Edvardsson}, {Enke},
  {Erdmann}, {Escolar}, {Espina}, {Evans}, {Eynard Bontemps}, {Fabre},
  {Fabrizio}, {Faigler}, {Falc{\~a}o}, {Farr{\`a}s Casas}, {Faye}, {Federici},
  {Fedorets}, {Fern{\'a}ndez-Hern{\'a}ndez}, {Fernique}, {Fienga}, {Figueras},
  {Filippi}, {Findeisen}, {Fonti}, {Fouesneau}, {Fraile}, {Fraser}, {Fuchs},
  {Furnell}, {Gai}, {Galleti}, {Galluccio}, {Garabato}, {Garc{\'\i}a-Sedano},
  {Gar{\'e}}, {Garofalo}, {Garralda}, {Gavras}, {Gerssen}, {Geyer}, {Gilmore},
  {Girona}, {Giuffrida}, {Gomes}, {Gonz{\'a}lez-Marcos},
  {Gonz{\'a}lez-N{\'u}{\~n}ez}, {Gonz{\'a}lez-Vidal}, {Granvik}, {Guerrier},
  {Guillout}, {Guiraud}, {G{\'u}rpide}, {Guti{\'e}rrez-S{\'a}nchez}, {Guy},
  {Haigron}, {Hatzidimitriou}, {Haywood}, {Heiter}, {Helmi}, {Hobbs},
  {Hofmann}, {Holl}, {Holland}, {Hunt}, {Hypki}, {Icardi}, {Irwin}, {Jevardat
  de Fombelle}, {Jofr{\'e}}, {Jonker}, {Jorissen}, {Julbe}, {Karampelas},
  {Kochoska}, {Kohley}, {Kolenberg}, {Kontizas}, {Koposov}, {Kordopatis},
  {Koubsky}, {Kowalczyk}, {Krone-Martins}, {Kudryashova}, {Kull}, {Bachchan},
  {Lacoste-Seris}, {Lanza}, {Lavigne}, {Le Poncin-Lafitte}, {Lebreton},
  {Lebzelter}, {Leccia}, {Leclerc}, {Lecoeur-Taibi}, {Lemaitre}, {Lenhardt},
  {Leroux}, {Liao}, {Licata}, {Lindstr{\o}m}, {Lister}, {Livanou}, {Lobel},
  {L{\"o}ffler}, {L{\'o}pez}, {Lopez-Lozano}, {Lorenz}, {Loureiro},
  {MacDonald}, {Magalh{\~a}es Fernandes}, {Managau}, {Mann}, {Mantelet},
  {Marchal}, {Marchant}, {Marconi}, {Marie}, {Marinoni}, {Marrese},
  {Marschalk{\'o}}, {Marshall}, {Mart{\'\i}n-Fleitas}, {Martino}, {Mary},
  {Matijevi{\v{c}}}, {Mazeh}, {McMillan}, {Messina}, {Mestre}, {Michalik},
  {Millar}, {Miranda}, {Molina}, {Molinaro}, {Molinaro}, {Moln{\'a}r},
  {Moniez}, {Montegriffo}, {Monteiro}, {Mor}, {Mora}, {Morbidelli}, {Morel},
  {Morgenthaler}, {Morley}, {Morris}, {Mulone}, {Muraveva}, {Musella},
  {Narbonne}, {Nelemans}, {Nicastro}, {Noval}, {Ord{\'e}novic},
  {Ordieres-Mer{\'e}}, {Osborne}, {Pagani}, {Pagano}, {Pailler}, {Palacin},
  {Palaversa}, {Parsons}, {Paulsen}, {Pecoraro}, {Pedrosa}, {Pentik{\"a}inen},
  {Pereira}, {Pichon}, {Piersimoni}, {Pineau}, {Plachy}, {Plum}, {Poujoulet},
  {Pr{\v{s}}a}, {Pulone}, {Ragaini}, {Rago}, {Rambaux}, {Ramos-Lerate},
  {Ranalli}, {Rauw}, {Read}, {Regibo}, {Renk}, {Reyl{\'e}}, {Ribeiro},
  {Rimoldini}, {Ripepi}, {Riva}, {Rixon}, {Roelens}, {Romero-G{\'o}mez},
  {Rowell}, {Royer}, {Rudolph}, {Ruiz-Dern}, {Sadowski}, {Sagrist{\`a}
  Sell{\'e}s}, {Sahlmann}, {Salgado}, {Salguero}, {Sarasso}, {Savietto},
  {Schnorhk}, {Schultheis}, {Sciacca}, {Segol}, {Segovia}, {Segransan},
  {Serpell}, {Shih}, {Smareglia}, {Smart}, {Smith}, {Solano}, {Solitro},
  {Sordo}, {Soria Nieto}, {Souchay}, {Spagna}, {Spoto}, {Stampa}, {Steele},
  {Steidelm{\"u}ller}, {Stephenson}, {Stoev}, {Suess}, {S{\"u}veges}, {Surdej},
  {Szabados}, {Szegedi-Elek}, {Tapiador}, {Taris}, {Tauran}, {Taylor},
  {Teixeira}, {Terrett}, {Tingley}, {Trager}, {Turon}, {Ulla}, {Utrilla},
  {Valentini}, {van Elteren}, {Van Hemelryck}, {van Leeuwen}, {Varadi},
  {Vecchiato}, {Veljanoski}, {Via}, {Vicente}, {Vogt}, {Voss}, {Votruba},
  {Voutsinas}, {Walmsley}, {Weiler}, {Weingrill}, {Werner}, {Wevers},
  {Whitehead}, {Wyrzykowski}, {Yoldas}, {{\v{Z}}erjal}, {Zucker}, {Zurbach},
  {Zwitter}, {Alecu}, {Allen}, {Allende Prieto}, {Amorim},
  {Anglada-Escud{\'e}}, {Arsenijevic}, {Azaz}, {Balm}, {Beck}, {Bernstein},
  {Bigot}, {Bijaoui}, {Blasco}, {Bonfigli}, {Bono}, {Boudreault}, {Bressan},
  {Brown}, {Brunet}, {Bunclark}, {Buonanno}, {Butkevich}, {Carret}, {Carrion},
  {Chemin}, {Ch{\'e}reau}, {Corcione}, {Darmigny}, {de Boer}, {de Teodoro}, {de
  Zeeuw}, {Delle Luche}, {Domingues}, {Dubath}, {Fodor}, {Fr{\'e}zouls},
  {Fries}, {Fustes}, {Fyfe}, {Gallardo}, {Gallegos}, {Gardiol}, {Gebran},
  {Gomboc}, {G{\'o}mez}, {Grux}, {Gueguen}, {Heyrovsky}, {Hoar}, {Iannicola},
  {Isasi Parache}, {Janotto}, {Joliet}, {Jonckheere}, {Keil}, {Kim},
  {Klagyivik}, {Klar}, {Knude}, {Kochukhov}, {Kolka}, {Kos}, {Kutka}, {Lainey},
  {LeBouquin}, {Liu}, {Loreggia}, {Makarov}, {Marseille}, {Martayan},
  {Martinez-Rubi}, {Massart}, {Meynadier}, {Mignot}, {Munari}, {Nguyen},
  {Nordlander}, {Ocvirk}, {O'Flaherty}, {Olias Sanz}, {Ortiz}, {Osorio},
  {Oszkiewicz}, {Ouzounis}, {Palmer}, {Park}, {Pasquato}, {Peltzer}, {Peralta},
  {P{\'e}turaud}, {Pieniluoma}, {Pigozzi}, {Poels}, {Prat}, {Prod'homme},
  {Raison}, {Rebordao}, {Risquez}, {Rocca-Volmerange}, {Rosen}, {Ruiz-Fuertes},
  {Russo}, {Sembay}, {Serraller Vizcaino}, {Short}, {Siebert}, {Silva},
  {Sinachopoulos}, {Slezak}, {Soffel}, {Sosnowska}, {Strai{\v{z}}ys}, {ter
  Linden}, {Terrell}, {Theil}, {Tiede}, {Troisi}, {Tsalmantza}, {Tur},
  {Vaccari}, {Vachier}, {Valles}, {Van Hamme}, {Veltz}, {Virtanen}, {Wallut},
  {Wichmann}, {Wilkinson}, {Ziaeepour}, \& {Zschocke}}]{2016A&A...595A...1G}
{Gaia Collaboration}, {Prusti}, T., {de Bruijne}, J.~H.~J., {et~al.} 2016,
  \aap, 595, A1, \dodoi{10.1051/0004-6361/201629272}

\bibitem[{{Gaia Collaboration} {et~al.}(2018){Gaia Collaboration}, {Brown},
  {Vallenari}, {Prusti}, {de Bruijne}, {Babusiaux}, {Bailer-Jones}, {Biermann},
  {Evans}, {Eyer}, {Jansen}, {Jordi}, {Klioner}, {Lammers}, {Lindegren},
  {Luri}, {Mignard}, {Panem}, {Pourbaix}, {Randich}, {Sartoretti}, {Siddiqui},
  {Soubiran}, {van Leeuwen}, {Walton}, {Arenou}, {Bastian}, {Cropper},
  {Drimmel}, {Katz}, {Lattanzi}, {Bakker}, {Cacciari}, {Casta{\~n}eda},
  {Chaoul}, {Cheek}, {De Angeli}, {Fabricius}, {Guerra}, {Holl}, {Masana},
  {Messineo}, {Mowlavi}, {Nienartowicz}, {Panuzzo}, {Portell}, {Riello},
  {Seabroke}, {Tanga}, {Th{\'e}venin}, {Gracia-Abril}, {Comoretto},
  {Garcia-Reinaldos}, {Teyssier}, {Altmann}, {Andrae}, {Audard},
  {Bellas-Velidis}, {Benson}, {Berthier}, {Blomme}, {Burgess}, {Busso},
  {Carry}, {Cellino}, {Clementini}, {Clotet}, {Creevey}, {Davidson}, {De
  Ridder}, {Delchambre}, {Dell'Oro}, {Ducourant},
  {Fern{\'a}ndez-Hern{\'a}ndez}, {Fouesneau}, {Fr{\'e}mat}, {Galluccio},
  {Garc{\'\i}a-Torres}, {Gonz{\'a}lez-N{\'u}{\~n}ez}, {Gonz{\'a}lez-Vidal},
  {Gosset}, {Guy}, {Halbwachs}, {Hambly}, {Harrison}, {Hern{\'a}ndez},
  {Hestroffer}, {Hodgkin}, {Hutton}, {Jasniewicz}, {Jean-Antoine-Piccolo},
  {Jordan}, {Korn}, {Krone-Martins}, {Lanzafame}, {Lebzelter}, {L{\"o}ffler},
  {Manteiga}, {Marrese}, {Mart{\'\i}n-Fleitas}, {Moitinho}, {Mora}, {Muinonen},
  {Osinde}, {Pancino}, {Pauwels}, {Petit}, {Recio-Blanco}, {Richards},
  {Rimoldini}, {Robin}, {Sarro}, {Siopis}, {Smith}, {Sozzetti}, {S{\"u}veges},
  {Torra}, {van Reeven}, {Abbas}, {Abreu Aramburu}, {Accart}, {Aerts},
  {Altavilla}, {{\'A}lvarez}, {Alvarez}, {Alves}, {Anderson}, {Andrei},
  {Anglada Varela}, {Antiche}, {Antoja}, {Arcay}, {Astraatmadja}, {Bach},
  {Baker}, {Balaguer-N{\'u}{\~n}ez}, {Balm}, {Barache}, {Barata}, {Barbato},
  {Barblan}, {Barklem}, {Barrado}, {Barros}, {Barstow}, {Bartholom{\'e}
  Mu{\~n}oz}, {Bassilana}, {Becciani}, {Bellazzini}, {Berihuete}, {Bertone},
  {Bianchi}, {Bienaym{\'e}}, {Blanco-Cuaresma}, {Boch}, {Boeche}, {Bombrun},
  {Borrachero}, {Bossini}, {Bouquillon}, {Bourda}, {Bragaglia}, {Bramante},
  {Breddels}, {Bressan}, {Brouillet}, {Br{\"u}semeister}, {Brugaletta},
  {Bucciarelli}, {Burlacu}, {Busonero}, {Butkevich}, {Buzzi}, {Caffau},
  {Cancelliere}, {Cannizzaro}, {Cantat-Gaudin}, {Carballo}, {Carlucci},
  {Carrasco}, {Casamiquela}, {Castellani}, {Castro-Ginard}, {Charlot},
  {Chemin}, {Chiavassa}, {Cocozza}, {Costigan}, {Cowell}, {Crifo}, {Crosta},
  {Crowley}, {Cuypers}, {Dafonte}, {Damerdji}, {Dapergolas}, {David}, {David},
  {de Laverny}, {De Luise}, {De March}, {de Martino}, {de Souza}, {de Torres},
  {Debosscher}, {del Pozo}, {Delbo}, {Delgado}, {Delgado}, {Di Matteo},
  {Diakite}, {Diener}, {Distefano}, {Dolding}, {Drazinos}, {Dur{\'a}n},
  {Edvardsson}, {Enke}, {Eriksson}, {Esquej}, {Eynard Bontemps}, {Fabre},
  {Fabrizio}, {Faigler}, {Falc{\~a}o}, {Farr{\`a}s Casas}, {Federici},
  {Fedorets}, {Fernique}, {Figueras}, {Filippi}, {Findeisen}, {Fonti},
  {Fraile}, {Fraser}, {Fr{\'e}zouls}, {Gai}, {Galleti}, {Garabato},
  {Garc{\'\i}a-Sedano}, {Garofalo}, {Garralda}, {Gavel}, {Gavras}, {Gerssen},
  {Geyer}, {Giacobbe}, {Gilmore}, {Girona}, {Giuffrida}, {Glass}, {Gomes},
  {Granvik}, {Gueguen}, {Guerrier}, {Guiraud}, {Guti{\'e}rrez-S{\'a}nchez},
  {Haigron}, {Hatzidimitriou}, {Hauser}, {Haywood}, {Heiter}, {Helmi}, {Heu},
  {Hilger}, {Hobbs}, {Hofmann}, {Holland}, {Huckle}, {Hypki}, {Icardi},
  {Jan{\ss}en}, {Jevardat de Fombelle}, {Jonker}, {Juh{\'a}sz}, {Julbe},
  {Karampelas}, {Kewley}, {Klar}, {Kochoska}, {Kohley}, {Kolenberg},
  {Kontizas}, {Kontizas}, {Koposov}, {Kordopatis}, {Kostrzewa-Rutkowska},
  {Koubsky}, {Lambert}, {Lanza}, {Lasne}, {Lavigne}, {Le Fustec}, {Le
  Poncin-Lafitte}, {Lebreton}, {Leccia}, {Leclerc}, {Lecoeur-Taibi},
  {Lenhardt}, {Leroux}, {Liao}, {Licata}, {Lindstr{\o}m}, {Lister}, {Livanou},
  {Lobel}, {L{\'o}pez}, {Managau}, {Mann}, {Mantelet}, {Marchal}, {Marchant},
  {Marconi}, {Marinoni}, {Marschalk{\'o}}, {Marshall}, {Martino}, {Marton},
  {Mary}, {Massari}, {Matijevi{\v{c}}}, {Mazeh}, {McMillan}, {Messina},
  {Michalik}, {Millar}, {Molina}, {Molinaro}, {Moln{\'a}r}, {Montegriffo},
  {Mor}, {Morbidelli}, {Morel}, {Morris}, {Mulone}, {Muraveva}, {Musella},
  {Nelemans}, {Nicastro}, {Noval}, {O'Mullane}, {Ord{\'e}novic},
  {Ord{\'o}{\~n}ez-Blanco}, {Osborne}, {Pagani}, {Pagano}, {Pailler},
  {Palacin}, {Palaversa}, {Panahi}, {Pawlak}, {Piersimoni}, {Pineau}, {Plachy},
  {Plum}, {Poggio}, {Poujoulet}, {Pr{\v{s}}a}, {Pulone}, {Racero}, {Ragaini},
  {Rambaux}, {Ramos-Lerate}, {Regibo}, {Reyl{\'e}}, {Riclet}, {Ripepi}, {Riva},
  {Rivard}, {Rixon}, {Roegiers}, {Roelens}, {Romero-G{\'o}mez}, {Rowell},
  {Royer}, {Ruiz-Dern}, {Sadowski}, {Sagrist{\`a} Sell{\'e}s}, {Sahlmann},
  {Salgado}, {Salguero}, {Sanna}, {Santana-Ros}, {Sarasso}, {Savietto},
  {Schultheis}, {Sciacca}, {Segol}, {Segovia}, {S{\'e}gransan}, {Shih},
  {Siltala}, {Silva}, {Smart}, {Smith}, {Solano}, {Solitro}, {Sordo}, {Soria
  Nieto}, {Souchay}, {Spagna}, {Spoto}, {Stampa}, {Steele},
  {Steidelm{\"u}ller}, {Stephenson}, {Stoev}, {Suess}, {Surdej}, {Szabados},
  {Szegedi-Elek}, {Tapiador}, {Taris}, {Tauran}, {Taylor}, {Teixeira},
  {Terrett}, {Teyssandier}, {Thuillot}, {Titarenko}, {Torra Clotet}, {Turon},
  {Ulla}, {Utrilla}, {Uzzi}, {Vaillant}, {Valentini}, {Valette}, {van Elteren},
  {Van Hemelryck}, {van Leeuwen}, {Vaschetto}, {Vecchiato}, {Veljanoski},
  {Viala}, {Vicente}, {Vogt}, {von Essen}, {Voss}, {Votruba}, {Voutsinas},
  {Walmsley}, {Weiler}, {Wertz}, {Wevers}, {Wyrzykowski}, {Yoldas},
  {{\v{Z}}erjal}, {Ziaeepour}, {Zorec}, {Zschocke}, {Zucker}, {Zurbach}, \&
  {Zwitter}}]{2018A&A...616A...1G}
{Gaia Collaboration}, {Brown}, A.~G.~A., {Vallenari}, A., {et~al.} 2018, \aap,
  616, A1, \dodoi{10.1051/0004-6361/201833051}

\bibitem[{{Ga{\l}an} {et~al.}(2019){Ga{\l}an}, {Miko{\l}ajewska}, {Monard},
  {I{\l}kiewicz}, {Pie{\'n}kowski}, \& {Gromadzki}}]{2019AcA....69...25G}
{Ga{\l}an}, C., {Miko{\l}ajewska}, J., {Monard}, B., {et~al.} 2019, \actaa, 69,
  25, \dodoi{10.32023/0001-5237/69.1.2}

\bibitem[{{Green} {et~al.}(2019){Green}, {Schlafly}, {Zucker}, {Speagle}, \&
  {Finkbeiner}}]{2019ApJ...887...93G}
{Green}, G.~M., {Schlafly}, E., {Zucker}, C., {Speagle}, J.~S., \&
  {Finkbeiner}, D. 2019, \apj, 887, 93, \dodoi{10.3847/1538-4357/ab5362}

\bibitem[{{Gustafsson} {et~al.}(2008){Gustafsson}, {Edvardsson}, {Eriksson},
  {J{\o}rgensen}, {Nordlund}, \& {Plez}}]{2008A&A...486..951G}
{Gustafsson}, B., {Edvardsson}, B., {Eriksson}, K., {et~al.} 2008, \aap, 486,
  951, \dodoi{10.1051/0004-6361:200809724}

\bibitem[{{Henize}(1960)}]{1960AJ.....65..491H}
{Henize}, K.~G. 1960, \aj, 65, 491, \dodoi{10.1086/108296}

\bibitem[{{Holtzman} {et~al.}(2015){Holtzman}, {Shetrone}, {Johnson}, {Allende
  Prieto}, {Anders}, {Andrews}, {Beers}, {Bizyaev}, {Blanton}, {Bovy},
  {Carrera}, {Chojnowski}, {Cunha}, {Eisenstein}, {Feuillet}, {Frinchaboy},
  {Galbraith-Frew}, {Garc{\'\i}a P{\'e}rez}, {Garc{\'\i}a-Hern{\'a}ndez},
  {Hasselquist}, {Hayden}, {Hearty}, {Ivans}, {Majewski}, {Martell},
  {Meszaros}, {Muna}, {Nidever}, {Nguyen}, {O'Connell}, {Pan}, {Pinsonneault},
  {Robin}, {Schiavon}, {Shane}, {Sobeck}, {Smith}, {Troup}, {Weinberg},
  {Wilson}, {Wood-Vasey}, {Zamora}, \& {Zasowski}}]{2015AJ....150..148H}
{Holtzman}, J.~A., {Shetrone}, M., {Johnson}, J.~A., {et~al.} 2015, \aj, 150,
  148, \dodoi{10.1088/0004-6256/150/5/148}

\bibitem[{{Huang} {et~al.}(2018){Huang}, {Liu}, {Chen}, {Zhang}, {Yuan},
  {Xiang}, {Wang}, \& {Tian}}]{2018AJ....156...90H}
{Huang}, Y., {Liu}, X.~W., {Chen}, B.~Q., {et~al.} 2018, \aj, 156, 90,
  \dodoi{10.3847/1538-3881/aacda5}

\bibitem[{{Iben} \& {Renzini}(1983)}]{1983ARA&A..21..271I}
{Iben}, I., J., \& {Renzini}, A. 1983, \araa, 21, 271,
  \dodoi{10.1146/annurev.aa.21.090183.001415}

\bibitem[{{Jorissen} \& {Mayor}(1988)}]{1988A&A...198..187J}
{Jorissen}, A., \& {Mayor}, M. 1988, \aap, 198, 187

\bibitem[{{Keenan}(1954)}]{1954ApJ...120..484K}
{Keenan}, P.~C. 1954, \apj, 120, 484, \dodoi{10.1086/145937}

\bibitem[{{Luo} {et~al.}(2012){Luo}, {Zhang}, {Zhao}, {Zhao}, {Cui}, {Li},
  {Chu}, {Shi}, {Wang}, {Zhang}, {Bai}, {Chen}, {Wang}, {Guo}, {Chen}, {Du},
  {Kong}, {Lei}, {Li}, {Song}, {Wu}, {Zhang}, {Zhou}, {Zuo}, {Du}, {He}, {Hou},
  {Dong}, {Li}, {Li}, {Li}, {Song}, {Tian}, {Wang}, {Wu}, {Yang}, {Yuan},
  {Cao}, {Chen}, {Chen}, {Chen}, {Chu}, {Feng}, {Gong}, {Gu}, {Hou}, {Huo},
  {Hu}, {Hu}, {Hu}, {Jia}, {Jiang}, {Jiang}, {Jiang}, {Jin}, {Li}, {Li}, {Li},
  {Li}, {Li}, {Liu}, {Liu}, {Liu}, {Lu}, {Lu}, {Luo}, {Mao}, {Men}, {Ni}, {Qi},
  {Qi}, {Shi}, {Su}, {Sun}, {Su}, {Tang}, {Tao}, {Tu}, {Wang}, {Wang}, {Wang},
  {Wang}, {Wang}, {Wang}, {Wang}, {Wang}, {Wang}, {Wang}, {Wang}, {Wang},
  {Wang}, {Wang}, {Wei}, {Xue}, {Xing}, {Xu}, {Xu}, {Xu}, {Yang}, {Yang},
  {Yao}, {Yu}, {Yuan}, {Zhai}, {Zhang}, {Zhang}, {Zhang}, {Zhang}, {Zhang},
  {Zhang}, {Zhao}, {Zhou}, {Zhu}, {Zhu}, \& {Zou}}]{2012RAA....12.1243L}
{Luo}, A.~L., {Zhang}, H.-T., {Zhao}, Y.-H., {et~al.} 2012, Research in
  Astronomy and Astrophysics, 12, 1243, \dodoi{10.1088/1674-4527/12/9/004}

\bibitem[{{MacConnell}(1979)}]{1979AAS...38..335M}
{MacConnell}, D.~J. 1979, \aaps, 38, 335

\bibitem[{{Majewski} {et~al.}(2017){Majewski}, {Schiavon}, {Frinchaboy},
  {Allende Prieto}, {Barkhouser}, {Bizyaev}, {Blank}, {Brunner}, {Burton},
  {Carrera}, {Chojnowski}, {Cunha}, {Epstein}, {Fitzgerald}, {Garc{\'\i}a
  P{\'e}rez}, {Hearty}, {Henderson}, {Holtzman}, {Johnson}, {Lam}, {Lawler},
  {Maseman}, {M{\'e}sz{\'a}ros}, {Nelson}, {Nguyen}, {Nidever}, {Pinsonneault},
  {Shetrone}, {Smee}, {Smith}, {Stolberg}, {Skrutskie}, {Walker}, {Wilson},
  {Zasowski}, {Anders}, {Basu}, {Beland}, {Blanton}, {Bovy}, {Brownstein},
  {Carlberg}, {Chaplin}, {Chiappini}, {Eisenstein}, {Elsworth}, {Feuillet},
  {Fleming}, {Galbraith-Frew}, {Garc{\'\i}a}, {Garc{\'\i}a-Hern{\'a}ndez},
  {Gillespie}, {Girardi}, {Gunn}, {Hasselquist}, {Hayden}, {Hekker}, {Ivans},
  {Kinemuchi}, {Klaene}, {Mahadevan}, {Mathur}, {Mosser}, {Muna}, {Munn},
  {Nichol}, {O'Connell}, {Parejko}, {Robin}, {Rocha-Pinto}, {Schultheis},
  {Serenelli}, {Shane}, {Silva Aguirre}, {Sobeck}, {Thompson}, {Troup},
  {Weinberg}, \& {Zamora}}]{2017AJ....154...94M}
{Majewski}, S.~R., {Schiavon}, R.~P., {Frinchaboy}, P.~M., {et~al.} 2017, \aj,
  154, 94, \dodoi{10.3847/1538-3881/aa784d}

\bibitem[{{Merrill}(1922)}]{1922ApJ....56..457M}
{Merrill}, P.~W. 1922, \apj, 56, 457, \dodoi{10.1086/142716}

\bibitem[{{Plez} {et~al.}(2003){Plez}, {van Eck}, {Jorissen}, {Edvardsson},
  {Eriksson}, \& {Gustafsson}}]{2003IAUS..210P..A2P}
{Plez}, B., {van Eck}, S., {Jorissen}, A., {et~al.} 2003, in Modelling of
  Stellar Atmospheres, ed. N.~{Piskunov}, W.~W. {Weiss}, \& D.~F. {Gray}, Vol.
  210, A2

\bibitem[{{Pr{\v{s}}a} {et~al.}(2016){Pr{\v{s}}a}, {Harmanec}, {Torres},
  {Mamajek}, {Asplund}, {Capitaine}, {Christensen-Dalsgaard}, {Depagne},
  {Haberreiter}, {Hekker}, {Hilton}, {Kopp}, {Kostov}, {Kurtz}, {Laskar},
  {Mason}, {Milone}, {Montgomery}, {Richards}, {Schmutz}, {Schou}, \&
  {Stewart}}]{2016AJ....152...41P}
{Pr{\v{s}}a}, A., {Harmanec}, P., {Torres}, G., {et~al.} 2016, \aj, 152, 41,
  \dodoi{10.3847/0004-6256/152/2/41}

\bibitem[{{Raskin} {et~al.}(2011){Raskin}, {van Winckel}, {Hensberge},
  {Jorissen}, {Lehmann}, {Waelkens}, {Avila}, {de Cuyper}, {Degroote},
  {Dubosson}, {Dumortier}, {Fr{\'e}mat}, {Laux}, {Michaud}, {Morren}, {Perez
  Padilla}, {Pessemier}, {Prins}, {Smolders}, {van Eck}, \&
  {Winkler}}]{2011A&A...526A..69R}
{Raskin}, G., {van Winckel}, H., {Hensberge}, H., {et~al.} 2011, \aap, 526,
  A69, \dodoi{10.1051/0004-6361/201015435}

\bibitem[{{Shetye} {et~al.}(2019){Shetye}, {Goriely}, {Siess}, {Van Eck},
  {Jorissen}, \& {Van Winckel}}]{2019A&A...625L...1S}
{Shetye}, S., {Goriely}, S., {Siess}, L., {et~al.} 2019, \aap, 625, L1,
  \dodoi{10.1051/0004-6361/201935296}

\bibitem[{{Shetye} {et~al.}(2020){Shetye}, {Van Eck}, {Goriely}, {Siess},
  {Jorissen}, {Escorza}, \& {Van Winckel}}]{2020A&A...635L...6S}
{Shetye}, S., {Van Eck}, S., {Goriely}, S., {et~al.} 2020, \aap, 635, L6,
  \dodoi{10.1051/0004-6361/202037481}

\bibitem[{{Shetye} {et~al.}(2018){Shetye}, {Van Eck}, {Jorissen}, {Van
  Winckel}, {Siess}, {Goriely}, {Escorza}, {Karinkuzhi}, \&
  {Plez}}]{2018A&A...620A.148S}
{Shetye}, S., {Van Eck}, S., {Jorissen}, A., {et~al.} 2018, \aap, 620, A148,
  \dodoi{10.1051/0004-6361/201833298}

\bibitem[{{Smith} \& {Lambert}(1990)}]{1990ApJS...72..387S}
{Smith}, V.~V., \& {Lambert}, D.~L. 1990, \apjs, 72, 387,
  \dodoi{10.1086/191421}

\bibitem[{{Stephenson}(1976)}]{1976PWSO...2...21S}
{Stephenson}, C.~B. 1976, Publications of the Warner \& Swasey Observatory, 2,
  2

\bibitem[{{Stephenson}(1984)}]{1984PWSO...3....1S}
---. 1984, Publications of the Warner \& Swasey Observatory

\bibitem[{{Stephenson}(1990)}]{1990AJ....100..569S}
---. 1990, \aj, 100, 569, \dodoi{10.1086/115540}

\bibitem[{{Su} \& {Cui}(2004)}]{2004ChJAA...4....1S}
{Su}, D.-Q., \& {Cui}, X.-Q. 2004, \cjaa, 4, 1, \dodoi{10.1088/1009-9271/4/1/1}

\bibitem[{{Suh}(2021)}]{2021ApJS..256...43S}
{Suh}, K.-W. 2021, \apjs, 256, 43, \dodoi{10.3847/1538-4365/ac1274}

\bibitem[{{Taylor}(2005)}]{2005ASPC..347...29T}
{Taylor}, M.~B. 2005, in Astronomical Society of the Pacific Conference Series,
  Vol. 347, Astronomical Data Analysis Software and Systems XIV, ed.
  P.~{Shopbell}, M.~{Britton}, \& R.~{Ebert}, 29

\bibitem[{{Tian} {et~al.}(2020){Tian}, {Liu}, {Yuan}, {Fang}, {Chen}, {Xiang},
  {Huang}, {Bi}, {Yang}, {Wu}, {Wang}, {Zhang}, {Huo}, {Yang}, {Liu}, {Guo}, \&
  {Zhang}}]{2020ApJS..249...22T}
{Tian}, Z., {Liu}, X., {Yuan}, H., {et~al.} 2020, \apjs, 249, 22,
  \dodoi{10.3847/1538-4365/ab9904}

\bibitem[{{Ting} {et~al.}(2018){Ting}, {Hawkins}, \&
  {Rix}}]{2018ApJ...858L...7T}
{Ting}, Y.-S., {Hawkins}, K., \& {Rix}, H.-W. 2018, \apjl, 858, L7,
  \dodoi{10.3847/2041-8213/aabf8e}

\bibitem[{{Van Eck} \& {Jorissen}(1999)}]{1999A&A...345..127V}
{Van Eck}, S., \& {Jorissen}, A. 1999, \aap, 345, 127.
\newblock \doarXiv{astro-ph/9903241}

\bibitem[{{Van Eck} \& {Jorissen}(2000)}]{2000A&A...360..196V}
---. 2000, \aap, 360, 196

\bibitem[{{Van Eck} {et~al.}(2000){Van Eck}, {Jorissen}, {Udry}, {Mayor},
  {Burki}, {Burnet}, \& {Catchpole}}]{2000A&AS..145...51V}
{Van Eck}, S., {Jorissen}, A., {Udry}, S., {et~al.} 2000, \aaps, 145, 51,
  \dodoi{10.1051/aas:2000349}

\bibitem[{{Van Eck} {et~al.}(2017){Van Eck}, {Neyskens}, {Jorissen}, {Plez},
  {Edvardsson}, {Eriksson}, {Gustafsson}, {J{\o}rgensen}, \&
  {Nordlund}}]{2017A&A...601A..10V}
{Van Eck}, S., {Neyskens}, P., {Jorissen}, A., {et~al.} 2017, \aap, 601, A10,
  \dodoi{10.1051/0004-6361/201525886}

\bibitem[{{Wang} {et~al.}(1996){Wang}, {Su}, {Chu}, {Cui}, \&
  {Wang}}]{1996ApOpt..35.5155W}
{Wang}, S.-G., {Su}, D.-Q., {Chu}, Y.-Q., {Cui}, X., \& {Wang}, Y.-N. 1996,
  \ao, 35, 5155, \dodoi{10.1364/AO.35.005155}

\bibitem[{{Wood} {et~al.}(1983){Wood}, {Bessell}, \&
  {Fox}}]{1983ApJ...272...99W}
{Wood}, P.~R., {Bessell}, M.~S., \& {Fox}, M.~W. 1983, \apj, 272, 99,
  \dodoi{10.1086/161265}

\bibitem[{{Wright} {et~al.}(2009){Wright}, {Barlow}, {Greimel}, {Drew},
  {Matsuura}, {Unruh}, \& {Zijlstra}}]{2009MNRAS.400.1413W}
{Wright}, N.~J., {Barlow}, M.~J., {Greimel}, R., {et~al.} 2009, \mnras, 400,
  1413, \dodoi{10.1111/j.1365-2966.2009.15536.x}

\bibitem[{{Wu} {et~al.}(2018){Wu}, {Xiang}, {Bi}, {Liu}, {Yu}, {Hon}, {Sharma},
  {Li}, {Huang}, {Liu}, {Zhang}, {Li}, {Ge}, {Tian}, {Zhang}, \&
  {Zhang}}]{2018MNRAS.475.3633W}
{Wu}, Y., {Xiang}, M., {Bi}, S., {et~al.} 2018, \mnras, 475, 3633,
  \dodoi{10.1093/mnras/stx3296}

\bibitem[{{Yang} \& {Jiang}(2011)}]{2011ApJ...727...53Y}
{Yang}, M., \& {Jiang}, B.~W. 2011, \apj, 727, 53,
  \dodoi{10.1088/0004-637X/727/1/53}

\bibitem[{{Yang} {et~al.}(2019){Yang}, {Bonanos}, {Jiang}, {Gao}, {Gavras},
  {Maravelias}, {Ren}, {Wang}, {Xue}, {Tramper}, {Spetsieri}, \&
  {Pouliasis}}]{2019A&A...629A..91Y}
{Yang}, M., {Bonanos}, A.~Z., {Jiang}, B.-W., {et~al.} 2019, \aap, 629, A91,
  \dodoi{10.1051/0004-6361/201935916}

\bibitem[{{Zhang} {et~al.}(2021){Zhang}, {Li}, {Yang}, {Xiong}, {Fu}, {Liu},
  {Tian}, {Li}, {Wang}, {Liang}, {Zhou}, {Zong}, {Yang}, {Liu}, \&
  {Hou}}]{2021ApJS..256...14Z}
{Zhang}, B., {Li}, J., {Yang}, F., {et~al.} 2021, \apjs, 256, 14,
  \dodoi{10.3847/1538-4365/ac0834}

\bibitem[{{Zhao} \& {Newberg}(2006)}]{2006astro.ph.12034Z}
{Zhao}, C., \& {Newberg}, H.~J. 2006, arXiv e-prints, astro.
\newblock \doarXiv{astro-ph/0612034}

\bibitem[{{Zhao} {et~al.}(2006){Zhao}, {Chen}, {Shi}, {Liang}, {Hou}, {Chen},
  {Zhang}, \& {Li}}]{2006ChJAA...6..265Z}
{Zhao}, G., {Chen}, Y.-Q., {Shi}, J.-R., {et~al.} 2006, \cjaa, 6, 265,
  \dodoi{10.1088/1009-9271/6/3/01}

\bibitem[{{Zhao} {et~al.}(2012){Zhao}, {Zhao}, {Chu}, {Jing}, \&
  {Deng}}]{2012RAA....12..723Z}
{Zhao}, G., {Zhao}, Y.-H., {Chu}, Y.-Q., {Jing}, Y.-P., \& {Deng}, L.-C. 2012,
  Research in Astronomy and Astrophysics, 12, 723,
  \dodoi{10.1088/1674-4527/12/7/002}

\bibitem[{{Zinn} {et~al.}(2019){Zinn}, {Pinsonneault}, {Huber}, \&
  {Stello}}]{2019ApJ...878..136Z}
{Zinn}, J.~C., {Pinsonneault}, M.~H., {Huber}, D., \& {Stello}, D. 2019, \apj,
  878, 136, \dodoi{10.3847/1538-4357/ab1f66}

\end{thebibliography}
\bibliographystyle{aasjournal}



\end{document}